\documentclass[aps,twocolumn,10pt,aps,amsmath,amssymb,nofootinbib,superscriptaddress,showkeys,showpacs]{revtex4-2}
\usepackage{epsfig} 
\usepackage{amsmath} 
\usepackage{graphicx}
\usepackage{color}
\usepackage{float}
\usepackage{soul,xcolor}
\usepackage[font=scriptsize]{caption}
\usepackage{braket}
\usepackage{bbold}
\usepackage{subfig}
\usepackage{braket}
\usepackage{titlesec}
\usepackage{placeins}
\usepackage{caption}
\usepackage{comment}
\usepackage{makecell}
\usepackage{rotating}
\usepackage{booktabs}
\usepackage{multirow}
\begin{document} 

\title{Searching for Pseudo-Dirac neutrinos from Astrophysical sources in IceCube data}

\author{Khushboo Dixit}
\email{kdixit@uj.ac.za}
\affiliation{Centre for Astro-Particle Physics (CAPP) and Department of Physics, University of Johannesburg, PO Box 524, Auckland Park 2006, South Africa}

\author{Luis Salvador Miranda}
\email{smiranda@ifsc.usp.br}
\affiliation{Instituto de Física de São Carlos, Universidade de São Paulo (USP) Av. Trabalhador São Carlense 400, São Carlos, São Paulo, Brazil}

\author{Soebur Razzaque}
\email{srazzaque@uj.ac.za}
\affiliation{Centre for Astro-Particle Physics (CAPP) and Department of Physics, University of Johannesburg, PO Box 524, Auckland Park 2006, South Africa}
\affiliation{Department of Physics, The George Washington University, Washington, DC 20052, USA}
\affiliation{National Institute for Theoretical and Computational Sciences (NITheCS), Private Bag X1, Matieland, South Africa}

\date{\today}

\begin{abstract}
We analyze IceCube public data from its IC86 configuration, namely PSTracks event selection, to search for pseudo-Dirac signatures in high-energy neutrinos from astrophysical sources. Neutrino flux from astrophysical sources is reduced in the pseudo-Dirac scenario due to the conversion of active-to-sterile neutrinos as compared to the neutrino oscillation scenario of only three active neutrinos over astrophysical distances. We fit IceCube data using astrophysical flux models for four point-like sources in both scenarios and constrain the active-sterile mass-square-difference in the absence of any evidence for the pseudo-Dirac scenario. We present the exclusion regions for the common mass-squared difference $\delta m^2$, inducing active-sterile oscillations, for all three neutrino flavors. This includes results from individual sources as well as from a stacking analysis that combines data from the four sources. Our findings indicate that the exclusion region is $\delta m^2 \in [2.1\times 10^{-21} - 2.0\times 10^{-16}]$ eV$^2$ with $\ge 90\%$  confidence level (CL) significance for neutrino energies ranging from 0.5 TeV to 1 PeV. When we extend the energy range down to 0.1 TeV, the exclusion region broadens to $\delta m^2 \in [1.1 \times 10^{-21} - 3.0\times 10^{-16}]$ eV$^2$ at $\ge 90\%$ CL.
\end{abstract}

\maketitle

\section{Introduction}
\label{intro}
The observation of high-energy neutrinos by IceCube from the direction of the gamma-ray blazar TXS 0506 +056 in coincident with a flaring event \cite{IceCube:2018dnn} provided the first hint of an astrophysical source of high-energy neutrinos. This hint was further strengthened by the detection of a neutrino flare from the direction of TXS 0506+056 prior to the flare detected in electromagnetic wavebands in coincident with a neutrino event \cite{IceCube:2018cha}. More recently, IceCube has established NGC 1068, an active galactic nuclei (AGN), to be a persistent source of high-energy neutrinos \cite{IceCube:2022der}. In fact it was pointed out early on \cite{Moharana:2015nxa} that NGC 1068 is in correlation with an ultrahigh cosmic-ray event detected by the Pierre Auger Observatory and a very high-energy neutrino event detected by IceCube.
Further time-integrated analysis of 10-year data, from 2008 to 2018, by IceCube Collaboration where it released its search results with aggregated neutrino emission observation from a list of 110 gamma-ray sources during this period \cite{IceCube:2021slf}. This excess of neutrino events is cognate mainly to four sources: NGC 1068, TXS 0506+056, PKS 1424+240 and GB6 J1542+6129, where the results show inconsistency with the background-only hypothesis in the Northern Hemisphere at the 3.3$\sigma$ level. Later on, IceCube found a global significance of 4.2$\sigma$ for NGC 1068 alone, while a binomial test combining the three sources NGC 1068, PKS 1424+240, and TXS 0506+056 yielded an overall (catalog-wide) significance of $\approx$ 3.7$\sigma$ local (3.4 $\sigma$ global) \cite{IceCube:2022der}. 
Recently, NGC~4151 has been identified as another neutrino source with 2.9$\sigma$ significance~\cite{IceCube:2024ayt}.

The observations of such persistent high energy astrophysical neutrino sources can act as natural neutrino beams and provide an opportunity to explore the presence of new physics effects in the neutrino sector. The flavor oscillation probabilities get averaged out over their frequency terms driven by the atmospheric and solar mass-squared differences associated to the mass eigenstates corresponding to the active neutrinos propagating over astrophysical distances. However, the advantage of observing possible effects of comparatively tiny mass-squared differences induced by some new physics effects can be observed in the neutrino fluxes from astrophysical sources. One such possibility is the pseudo-Dirac nature of neutrinos. In this scenario, neutrinos still exhibit the characteristics of a Dirac particle, while having a tiny Majorana mass as well. This is also known as the soft lepton number violating case as the small Majorana mass induces very small but nonzero lepton-number violation. It makes this scenario interesting to explore as it can shed some light on the nature of neutrino mass. The details of the quasi-Majorana mass scheme have been delineated with details in \cite{Wolfenstein:1981kw,Petcov:1982ya,Bilenky:1987ty,Joshipura:2000ts,Kobayashi:2000md,PhysRevD.80.073007,PhysRevD.97.095008}. 

The pseudo-Dirac scenario speculates the existence of extra sterile flavor states along with the active ones. However, due to the small Majorana masses, there is a minor splitting between the active and sterile states, which is negligible in general but can be observed efficiently if the travel distance of neutrinos is significantly large and/or neutrinos have very small energy. So far, the constraints on the active-sterile mass splitting have been obtained from atmospheric neutrinos as $\lesssim 10^{-4}~eV^2$ \cite{Beacom:2003eu}, from solar neutrinos $\lesssim 10^{-11}~eV^2$ \cite{PhysRevD.80.073007,Ansarifard:2022kvy} as well as from supernovae neutrinos that provide a narrow valid range of $\delta m^2 \sim [2.55, 3.01] \times 10^{-20}~eV^2$  \cite{DeGouvea:2020ang,Martinez-Soler:2021unz}. Moreover, some weaker constraints are also available from the LHC data \cite{Das:2014jxa}.

A favorable situation to probe pseudo-Dirac neutrinos can be provided by astrophysical neutrinos where neutrinos travel the distance up to 100s to 1000s of Mpc. In some previous works, constraint on the tiny mass splitting driving the active-sterile neutrino oscillations has been obtained in case of high energy astrophysical neutrinos such as using the IceCube data regarding the NGC 1068 source \cite{Rink:2022nvw}. At the same time, a more detailed analysis was done in terms of the sensitivity of the IceCube data from PKS 1424+240 and TXS 0506+056 sources along with NGC 1068 for this scenario in Ref.~\cite{Carloni:2022cqz}. Some other bounds have been established recently using the astrophysical neutrino data in this line. For instance, a bound has been obtained on $\delta m^2_3 \leq 10^{-12}$~eV$^2$ for the third pseudo-Dirac mass eigenstates using the astrophysical neutrino flavor data \cite{Fong:2024msb}. Additionally, the range of $\delta m^2$ $\in [5\times 10^{-19}, 8\times 10^{-19}]$eV$^2$ has been excluded at $3\sigma$ CL using the diffuse astrophysical neutrino data \cite{Carloni:2025dhv}. Another interesting work explored the matter effect on active-sterile oscillations for pseudo-Dirac neutrinos induced by the cosmic neutrino background \cite{Dev:2024yrg}. 

In this work, we use the IceCube public data from its IC86 configuration, which is named as PSTracks event selection and put constraint on the active-sterile mass squared difference from individual sources: NGC 1068, NGC~4151, PKS 1424+240 and TXS 0506+056. Also, we perform a stacking analysis and provide the constraint on the same parameters. We have generalized this analysis by keeping the spectral index for the source flux and source event counts as free parameters while fitting data both in the standard oscillation scenario and in the pseudo-Dirac case. 
Also, we provide our results for different energy ranges considered for the neutrino data. In Sec.~\ref{sec:1}, we discuss the phenomena of pseudo-Dirac neutrino oscillations and provide the expressions for survival and transition probabilities. 

In Sec.~\ref{sec:ICdata} we discuss the details of IceCube PSTrack data selection and source properties in Sec.~\ref{subsec:sources}, followed by the astrophysical neutrino fluxes observed on the earth in Section.~\ref{subsec:2}. 
Then we provide the statistical methodology for data analysis used to obtain the constraints on active-sterile mass splitting in Sec.~\ref{sec:analysis}. Finally, we present our results in Sec.~\ref{sec:results}, and discuss our results and conclude in Sec.~\ref{sec:conclusions}.

\section{Pseudo-Dirac scenario}
\label{sec:1}
The phenomena of neutrino oscillations indicate that neutrinos are not massless. Since then it has induced immense interest to find out the mechanism to produce neutrino masses.
One way to explain neutrino masses is to extend the Standard Model (SM) to include three additional sterile neutrinos where at least two of them have nonzero Majorana masses. In this scenario, the mass matrix takes the form as follows
\begin{equation}
	M = \begin{pmatrix}
		M_{L}  & M_{D}^T\\
		M_{D}  & M_{R}^{\ast}
	\end{pmatrix}
\end{equation}
where $M_{D}$, $M_{L}$ and $M_{R}$ are the Dirac, the left-handed Majorana and right-handed Majorana mass terms, respectively, in terms of 3$\times$3 matrices for three generations. The Dirac masses are consequences of the so-called Yukawa couplings, while on the other hand, the nonzero Majorana masses induce lepton number violation. An intermediate case of soft Lepton number violation is also possible with a very small value of Majorana mass term compared to the Dirac mass, $i.e.,$ $M_{L,R} \ll M_{D}$. In this case, the given mass matrix can be diagonalized using a 6$\times$6 block-diagonal unitary matrix that consists of the traditional PMNS matrix and a 3$\times$3 mixing matrix associated with the right-handed (sterile) neutrinos. 
This scenario is known as pseudo-Dirac or quasi-Dirac neutrinos \cite{Wolfenstein:1981kw,Petcov:1982ya,Bilenky:1987ty,Joshipura:2000ts,Kobayashi:2000md,PhysRevD.80.073007,PhysRevD.97.095008}. The mixing between active-sterile pairs associated to each flavor is almost maximum due to very small Majorana mass terms. Hence, in the pseudo-Dirac scenario, each neutrino flavor state becomes the superposition of two almost non-degenerate mass eigenstates as given below
\begin{equation}
	\nu_{\alpha L} = U_{\alpha j}\frac{(\nu_j^+ + i \nu_j^-)}{\sqrt{2}} .
\end{equation}
Here, $\nu_j^{\pm}$ are mass eigenstates with corresponding masses $m^2_{j,\pm} = m^2_{j}\pm \delta m_j^2$, where $\delta m_j^2$ is the mass-squared difference that drives oscillations between the active and sterile states. In the case of zero $\delta m_j^2$, $m_j^2$ represent masses associated with the mass eigenstates in the usual active three-flavor oscillation framework. 
The mixing matrix for such active-sterile mixing for three flavor oscillation scenario can be expressed as \cite{Kobayashi:2000md}
\vspace{3mm}
\noindent

\begin{widetext}
	\small
	\begin{align}
		V &= \begin{pmatrix}
			U_{\rm PMNS}  & 0\\
			0     & U_{\rm R}
		\end{pmatrix} \begin{pmatrix}
			\frac{1}{\sqrt{2}}  &0  &0  &\frac{i}{\sqrt{2}}  &0  &0\\
			0  &\frac{1}{\sqrt{2}}  &0  &0  &\frac{i}{\sqrt{2}}  &0\\
			0  &0  &\frac{1}{\sqrt{2}}  &0  &0  &\frac{i}{\sqrt{2}}\\
			\frac{1}{\sqrt{2}}e^{-i \phi_1}  &0  &0  &-\frac{i}{\sqrt{2}}e^{-i \phi_1}  &0  &0\\
			0  &\frac{1}{\sqrt{2}}e^{-i \phi_2}  &0  &0  &-\frac{i}{\sqrt{2}}e^{-i \phi_2}  &0\\
			0  &0  &\frac{1}{\sqrt{2}}e^{-i \phi_3}  &0  &0  &-\frac{i}{\sqrt{2}}e^{-i \phi_3}
		\end{pmatrix},
		\label{eq:mixing_matrix}
	\end{align}
	\normalsize
\end{widetext}
\noindent 
and the Hamiltonian in mass basis is given as 
\begin{equation*}
	M = diag(m_{1a},m_{2a},m_{3a},m_{1s},m_{2s},m_{3s}).
\end{equation*}
In Eq.~(\ref{eq:mixing_matrix}), $U_{\rm R}$ is the 3$\times$3 unitary matrix driving mixing between the sterile states. Interestingly, the active flavor transition probabilities do not depend on the elements of $U_{\rm R}$.

\begin{figure}[h!]
	\centering
	\includegraphics[width=.5\textwidth]{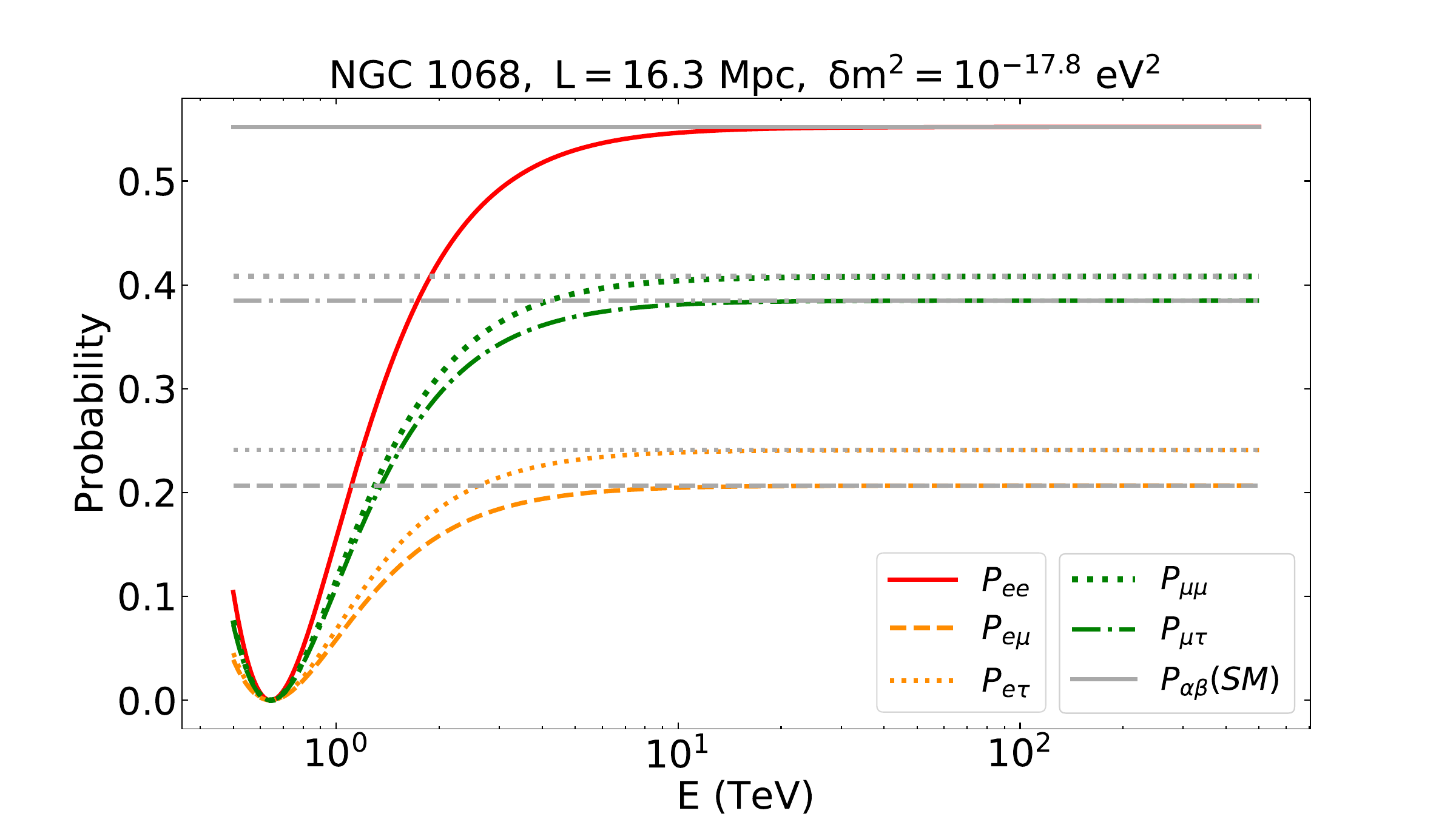}
	\includegraphics[width=.5\textwidth]{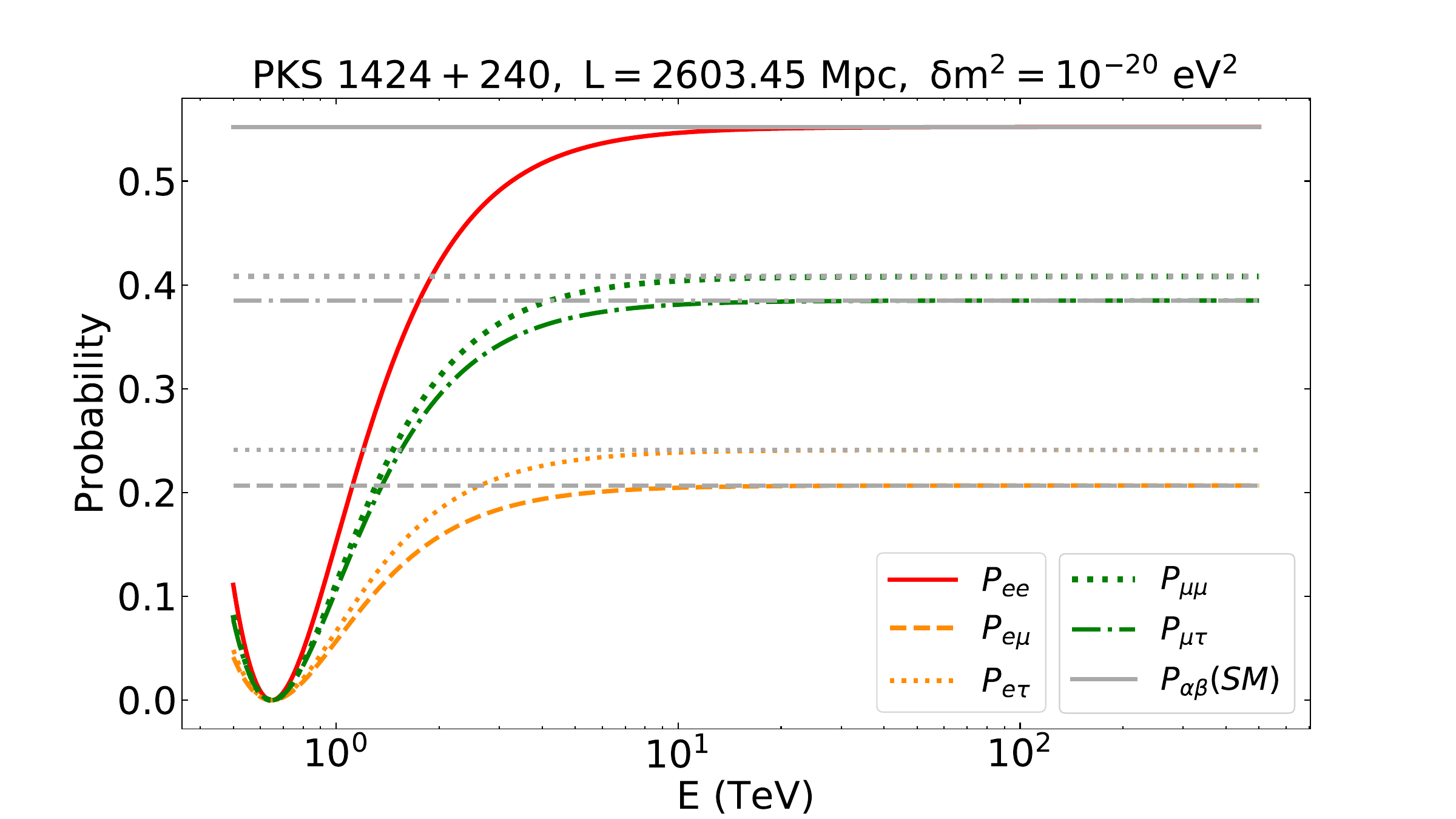}
	\includegraphics[width=.5\textwidth]{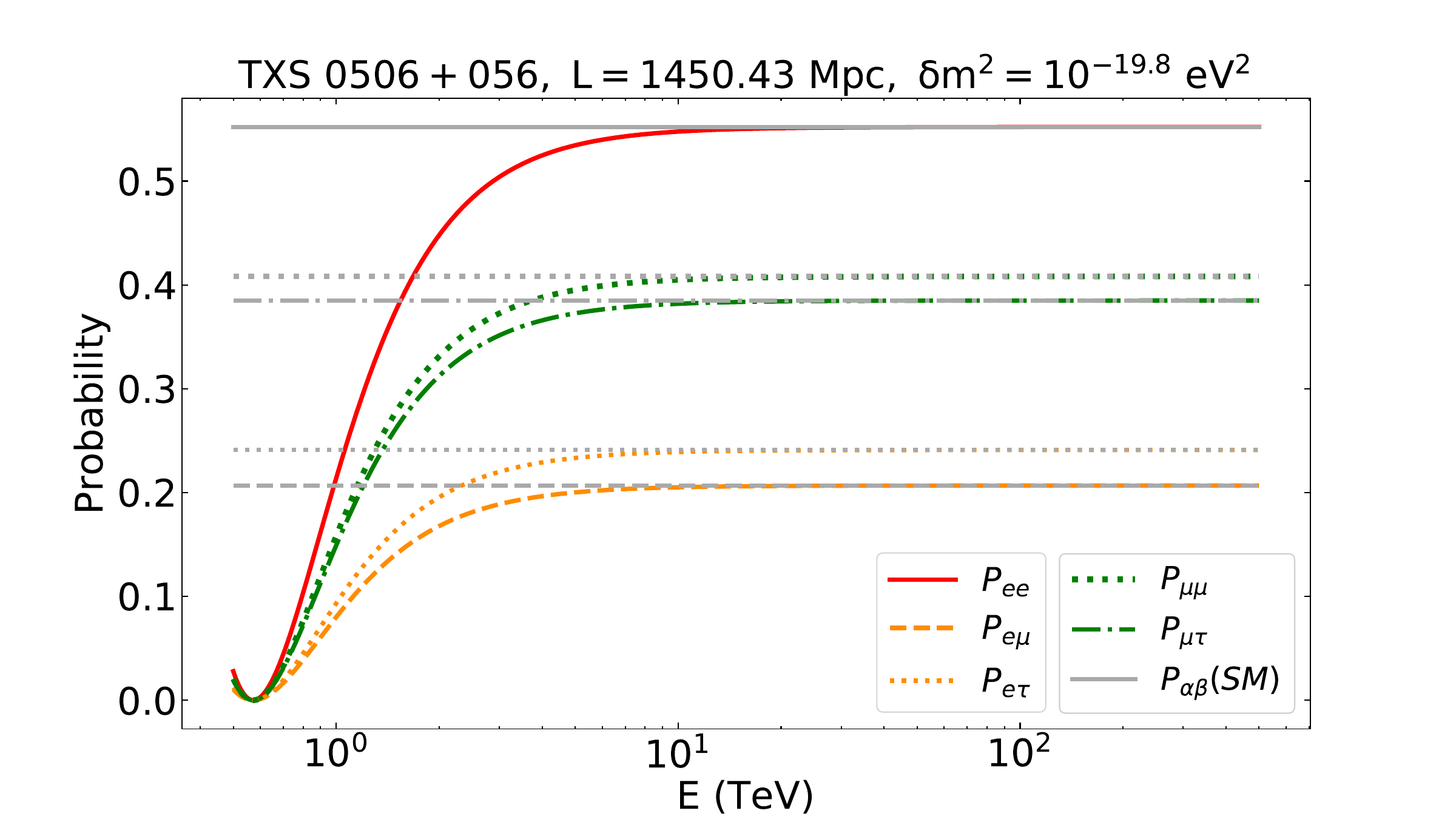}
	\includegraphics[width=.5\textwidth]{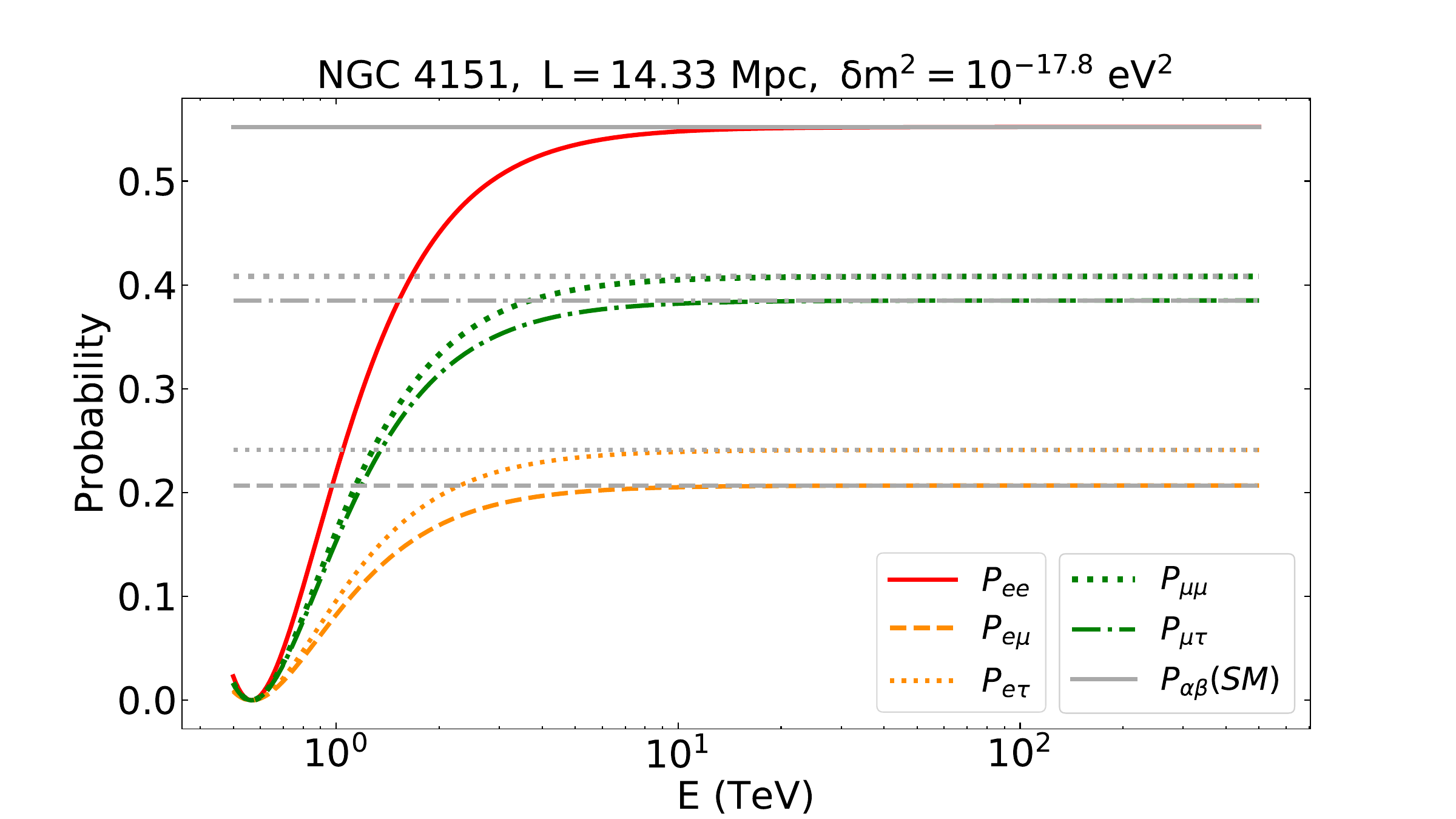}
	\caption{Oscillation probabilities for the three active neutrinos in the pseudo-Dirac scenario with respect to the neutrino energy $E_{\nu}$ for NGC 1068, PKS 1424+240, TXS 0506+056 and NGC 4151 for different $\delta m^2$ values. 
	}
	\label{fig:probabilities_En} 
\end{figure}

\begin{figure}[h!]
	\centering
	\includegraphics[width=.5\textwidth]{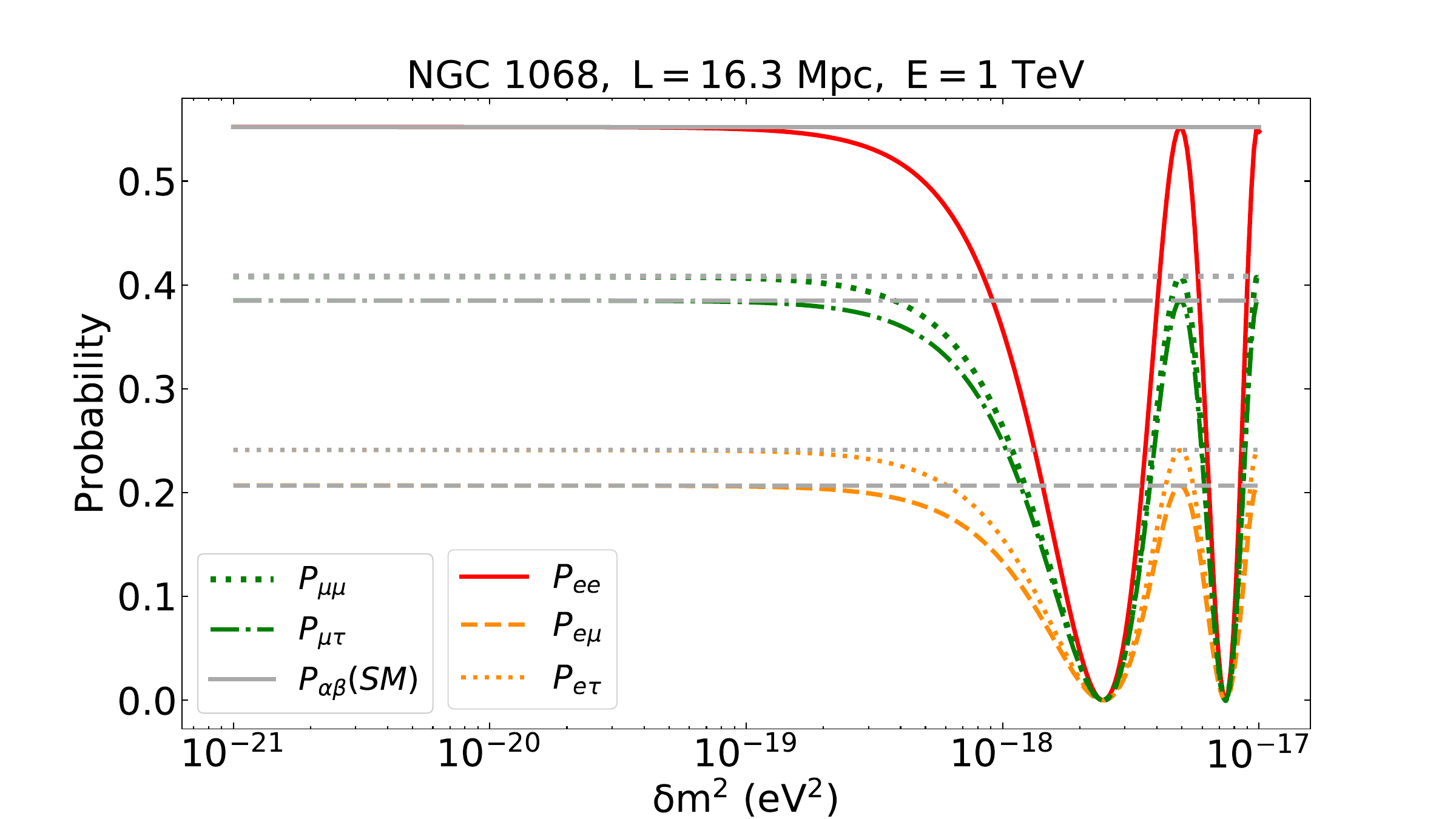}
	\includegraphics[width=.5\textwidth]{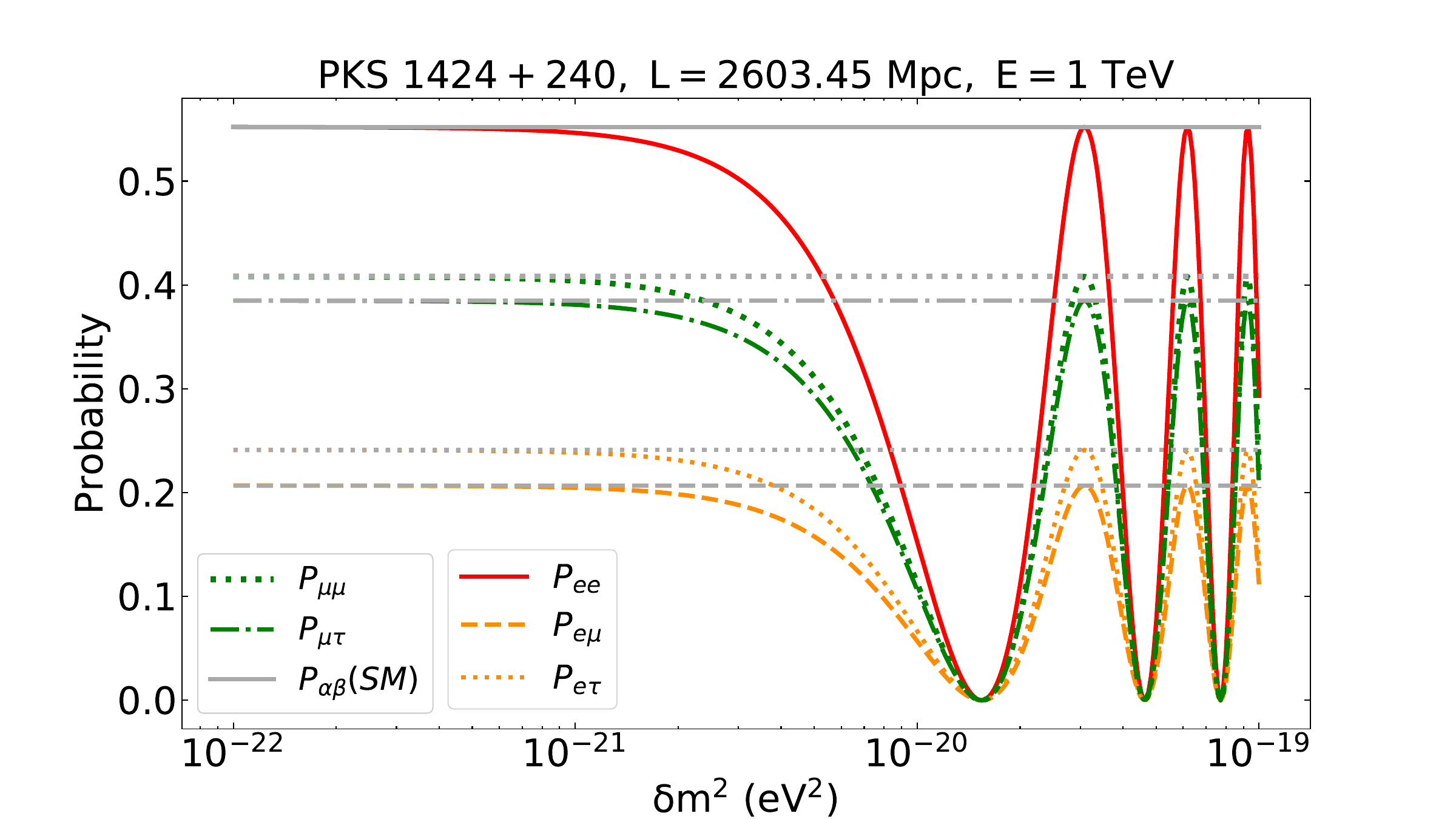}
	\includegraphics[width=.5\textwidth]{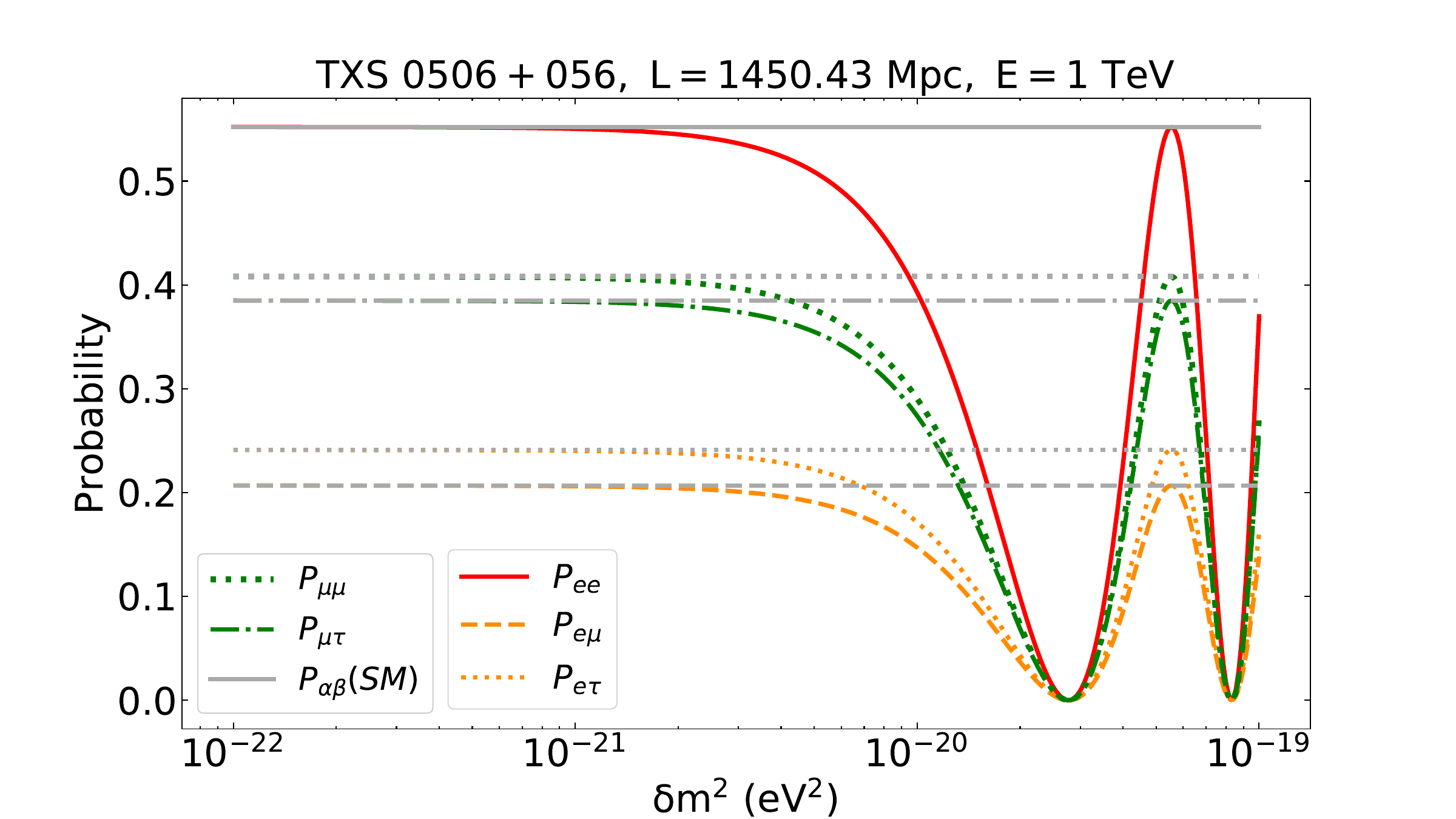} 
	\includegraphics[width=.5\textwidth]{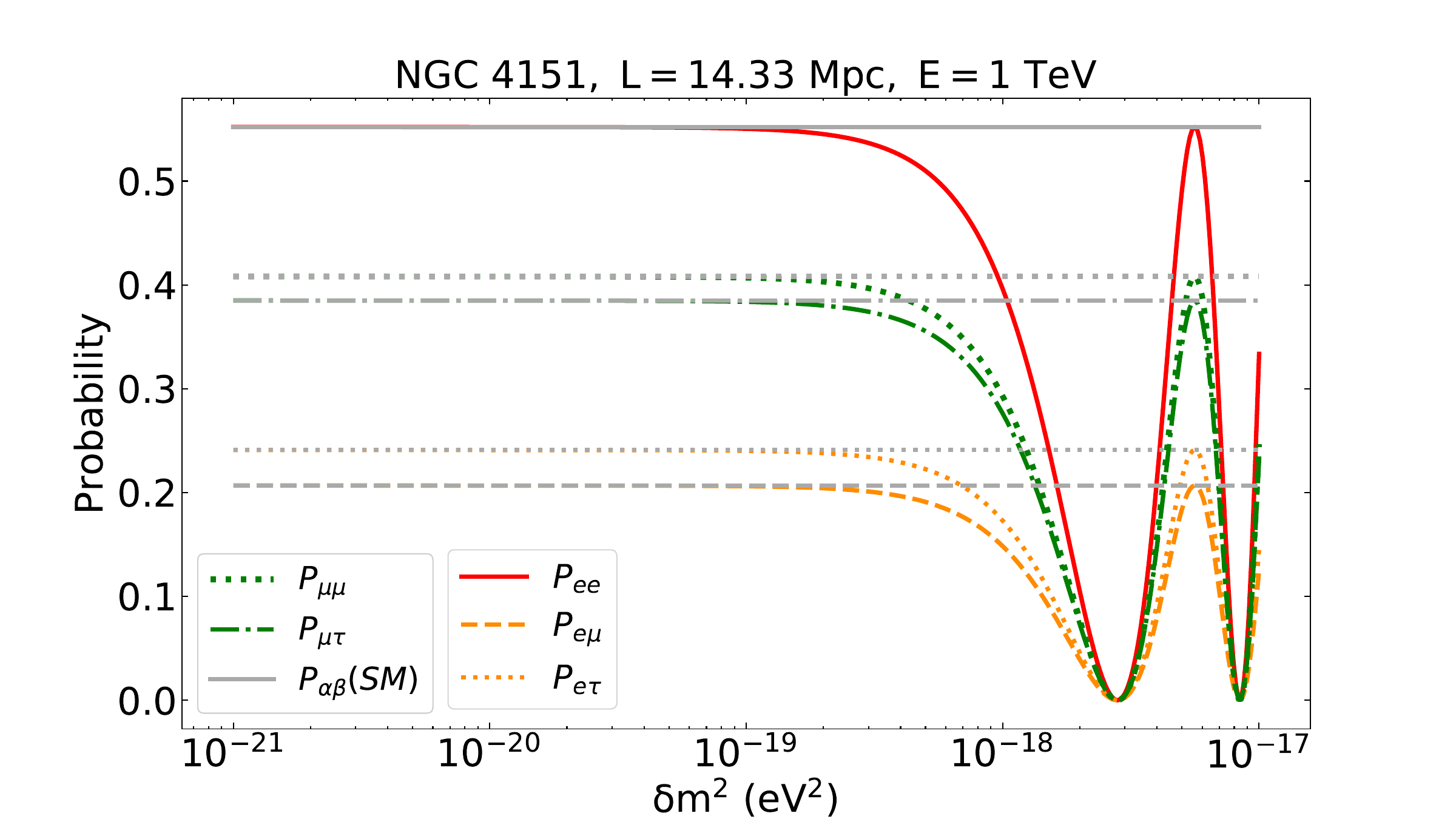}
	\caption{Variation of oscillation probabilities for the three active neutrinos in the pseudo-Dirac scenario with respect to the $\delta m_i^2$ for $E_{\nu} = 1$ TeV, in the context of NGC 1068, PKS 1424+240, TXS 0506+056 and NGC 4151. We considered here $\delta m_1^2 = \delta m_2^2 = \delta m_3^2 = \delta m^2$. Different curves represent different survival and transition probabilities. The corresponding SM probabilities are shown as the gray lines, e.g., $P_{ee}^{\rm SM}$ is shown as the gray solid line. 
	}
	\label{fig:probabilities_dmsqr} 
\end{figure}

The survival and transition probabilities of active flavors propagating over astrophysical distances in the pseudo-Dirac scenario take the forms as follows \cite{Kobayashi:2000md,Esmaili:2012ac}
\begin{align}
	P_{\alpha\alpha} &= \sum_{i=1,2,3} |U_{\alpha i}|^4 \cos ^2\left(\frac{\delta m_i^2 L}{4 E_{\nu}}\right), \nonumber \\
	P_{\alpha\beta} &= \sum_{i=1,2,3} |U_{\alpha i}|^2 |U_{\beta i}|^2 \cos ^2\left(\frac{\delta m_i^2 L}{4 E_{\nu}}\right),
	\label{eq:pD_probs}
\end{align}
where $\alpha, \beta = e,\mu,\tau$. Note that $P_{\alpha\alpha}$ and $P_{\alpha\beta}$ reduce to the SM case for $\delta m^2_i = 0$.
Note also that in the special case when $\delta m_1^2 = \delta m_2^2 = \delta m_3^2 = \delta m^2$, the transition probabilities for pseudo-Dirac neutrinos in Eq.~(\ref{eq:pD_probs}) take the form 
\begin{equation}\label{Probequalmasssqr}
	P_{\alpha \beta} = P_{\alpha \beta}^{\rm SM} ~\cos^2\left(\frac{\delta m^2 L}{4 E}\right) ,
\end{equation}  
where $P_{\alpha\beta}^{\rm SM} = \sum_{i=1,2,3} |U_{\alpha i}|^2 |U_{\beta i}|^2$ represent the flavor transition probabilities for the standard case of Dirac neutrinos which is averaged over astrophysical distances.
The probabilities in Eq.~(\ref{eq:pD_probs}) are plotted with respect to the mass squared difference $\delta m_i^2 \equiv \delta m^2$, are shown in Fig.~\ref{fig:probabilities_dmsqr} for the distance of the sources NGC 1068, PKS 1424+240, TXS 0506+056 and NGC~4151 from top to bottom. Here, sensitivity to the smallest values of $\delta m^2$ can be seen for PKS 1424+240 which can be attributed to the longest distance observed of this source among all four mentioned sources here. Moreover, the variation of these flavor transition probabilities can also be seen in Fig.~\ref{fig:probabilities_En} for different sources and corresponding $\delta m^2$ values of different orders of magnitudes. It can be seen that the effects of these tiny $\delta m_i^2$ values are significant at lower neutrino energy ranges.

\section{Astrophysical neutrino data and sources}
\subsection{IceCube PSTracks data}
\label{sec:ICdata}
The IceCube Collaboration in 2018 released a public data set of neutrino track events from its 40-string configuration (IC40) starting from 2008 throughout the full 86-string configuration (IC86).  This event selection is called PSTracks v3 and it provides an invaluable opportunity to implement correlation studies with the Galactic and extragalactic sources, due to a very good angular resolution of the data \cite{IceCube:2021xar,IceCube:2019cia}.
The public data set contains one-year of data for each of the configurations IC40, IC59 and IC79 and seven years of data for the IC86 configuration. The data release also includes files of binned instrument response functions and effective areas.

In order to perform a consistent analysis we decided to use the track neutrino events from the IC86 configuration only. Northern and Southern hemispheres are split at $-5$ degrees in declination, and for values larger than 81 degrees studies are not reliable. Therefore we limit this study to the declination range $[-5^\circ, 81^\circ]$, containing a total of 601,163 muon neutrino or track events collected over seven years in the IC86 configuration.

\subsection{Astrophysical neutrino sources}
\label{subsec:sources}

\begin{table*}[t]
\caption{List of highest-significance point sources in 2008--2018 IceCube data release~\cite{IceCube:2021slf} along with their positions in the sky and distance. We have also included in this work the source NGC 4151 that has been recently reported by IceCube collaboration in Ref.~\cite{IceCube:2024ayt}.}
\label{tab:sources}
\begin{ruledtabular}
\begin{tabular}{lcccc}
Name & RA (Deg) & Dec (Deg) & Redshift & Distance (Mpc) \\
NGC 1068       & 40.669629  & -0.013281 & 0.00379      & 16.3 \\
TXS 0506+056   & 77.358185  & 5.693148  & 0.3365       & 1450.43 \\
PKS 1424+240   & 216.751632 & 23.8      & 0.16 (0.604) & 689.65 (2603.45) \\
GB6 J1542+6129 & 235.737265 & 61.498707 & 0.34--1.76   & 1465.52--4896.5 \\
NGC 4151       & 182.635755 & 39.405849 & 0.003326     & 14.336 \\
\end{tabular}
\end{ruledtabular}
\end{table*}

In its analysis of ten years of data, the IceCube Collaboration has found four significant point sources in the sky at 3.3$\sigma$ level \cite{IceCube:2021slf}. A fifth source, NGC~4151, has been announced more recently with 2.9$\sigma$ significance~\cite{IceCube:2024ayt}. These neutrino sources are aligned with known astrophysical objects as listed in Table~\ref{tab:sources} below. The coordinates of the sources were obtained from the NED\footnote{The NASA/IPAC Extragalactic Database (NED) is funded by the National Aeronautics and Space Administration and operated by the California Institute of Technology.} There are two possible redshift values for PKS~1424+240, $z=0.16$ \cite{MAGIC:2010ovt} and $0.604$ \cite{Furniss:2013roa,Paiano:2017pol}. There is no consensus on the redshift of GB6~J1542+6129. Based on the spectroscopy of gamma-ray BL lac sample, its redshift is in the range of 0.34-1.76~\cite{Shaw:2013pp,Padovani:2022wjk}. As a result, we have ignored this source in our analysis. We calculate the comoving distance for $H_0 = 69.6$~km~sec$^{-1}$~Mpc$^{-1}$, $\Omega_{\rm M} = 0.286$ and $\Omega_{\Lambda} = 0.714$ using Ref.~\cite{Wright:2006up}.

\subsection{Source neutrino flux}
\label{subsec:2}
In our analysis, we consider pion-decay scenario, which are produced via $pp$ and/or $p\gamma$ interactions, for neutrino production: $\pi^+ \to \mu^+ \nu_\mu \to e^+ \nu_e {\bar\nu}_\mu \nu_\mu$ for the initial neutrino fluxes. The ratio of $\nu_\mu$ to $\nu_e$ fluxes and of $\nu_\mu$ to ${\bar \nu}_\mu$ fluxes can be estimated using the decay properties of $\pi^+\to \mu^+ \nu_\mu$ and $\mu^+\to e^+ \nu_e {\bar\nu}_\mu \nu_\mu$, as well as on the spectrum of the primary protons \cite{Lipari:1993hd,Razzaque:2013dsa}. For a primary proton spectrum $dN/dE \propto E^{-p}$,  
the initial neutrino fluxes also follow power laws and are approximately, 
\begin{eqnarray}
	\Phi_{\nu_\mu + {\bar \nu}_\mu}^0 \approx \Phi_{\nu_e + {\bar \nu}_e}^0 = \Phi^0 \left( 
	\frac{E_\nu}{{\rm TeV}} \right)^{-\gamma}\,,
	\label{eq:source_fluxes}
\end{eqnarray}
where we have assumed simple power-law fluxes with $\gamma$ representing the spectral index that indicates the slope of the spectrum and $\Phi^0$ is the normalization. These ratios remain approximately the same in case of no significant energy losses by pions and muons before decaying \cite{Rachen:1998fd}.
Hence, we use the source flux of $\nu_\mu$ or ${\bar \nu}_\mu$ as
\begin{eqnarray}
	\Phi_{\nu_\mu}^{\rm src} = x P_{e\mu}\Phi_{\nu_{e}}^0+(1-x)P_{\mu\mu}\Phi^0_{\nu_{\mu}} ,
	\label{eq:source_fluxes2}
\end{eqnarray}
after neutrino propagation through space and Earth. We assume that the neutrino production mechanism is $\pi$-decay, which results in an initial flavor ratio of $\nu_e:\nu_\mu:\nu_\tau = 1:2:0$ at the source, represented as $x = 1/3$. In case of $n$-decay and $\mu$-damped $\pi$-decay channels, $x$ takes the values of $1$ and $0$, respectively. Our results, however, do not depend strongly on the choice of $x$ since we essentially vary $\Phi^0$ and $\gamma$. 

\section{Data analysis}
\label{sec:analysis}
We calculate the number of neutrino events both from the astrophysical point sources and from the backgrounds in this section and perform statistical analysis to constrain the model parameters, including the mass-squared-difference $\delta m_i^2$ between the active and sterile neutrinos in the pseudo-Dirac scenario of neutrinos. 

We calculate the number of neutrino events both from the astrophysical point sources and from the backgrounds in this section and perform statistical analysis to constrain the model parameters, including the mass-squared-difference $\delta m_i^2$ between the active and sterile neutrinos in the pseudo-Dirac scenario of neutrinos.  

\subsection{Signal and background events}
\label{subsec:events}
We calculate the number of signal $\nu_\mu + {\bar\nu}_\mu$ events from an astrophysical source in an energy bin $E_k-E_{k+1}$ over the detector lifetime $T$, using the source flux in Eqs.~(\ref{eq:source_fluxes}) and (\ref{eq:source_fluxes2}) as 
\begin{align}
	n_{s,k} = T \int d\Omega \int_{E_k}^{E_{k+1}} dE_\nu\, &A_\nu^{\rm eff} (E_\nu, \Omega) 
	\left[
	\phi_{\nu_\mu}^{\rm src} (E_\nu; \delta m^2, \phi^0, \gamma)\right. \nonumber \\
	&\left.+
	\phi_{{\bar \nu}_\mu}^{\rm src} (E_\nu; \delta m^2, \phi^0, \gamma)
	\right],
	\label{eq:signal_evts}
\end{align}
where $A_\nu^{\rm eff}$ is the neutrino effective area of the detector. We calculate the background events from the atmospheric and from a diffuse astrophysical flux as,
\begin{align}
	n_{b,k} = T \int d\Omega &\int_{E_k}^{E_{k+1}} dE_\nu\, A_\nu^{\rm eff} (E_\nu, \Omega) 
	\left[
	\phi_{\nu_\mu}^{\rm atm} (E_\nu, \Omega)\right. \nonumber \\
	&\left.+ 
	\phi_{\nu_\mu}^{\rm ast} (E_\nu, \Omega)
	\right]  + {\rm antineutrino~events.}
	\label{eq:background_evts}
\end{align}
We use the conventional atmospheric neutrino fluxes from the Honda et al.\ model \cite{Honda:2015fha} and the prompt atmospheric neutrino flux component from the Enberg \& Reno model \cite{Enberg:2008te}. For the diffuse astrophysical neutrino background, we use a power-law of the form
\begin{eqnarray}
	\phi_{\nu_\mu}^{\rm ast}  = 
	\phi_{\rm ast} \left( \frac{E_{\nu}}{100~{\rm TeV}} \right) ^{-\gamma_{\rm ast}} ,
	\label{eq:diff_ast_flux}
\end{eqnarray}
with $\phi_{\rm ast}=1.44\times 10^{-18} ~\rm{GeV^{-1}cm^{-2}s^{-1}sr^{-1}}$ and $\gamma_{\rm{ast}}=2.28$ from \cite{Stettner:2019tok} kept fixed.
We use energy resolution $\log_{10} (\Delta E/E)=0.3$ 
and angular resolutions 
given in the PSTracks data. We considered in this work different ranges of energy to perform the analysis. The results presented here are for 0.5 TeV - 1 PeV energy range. We also provide the results for 0.1 TeV - 1 PeV range in the Appendix. The results from both these energy-ranges are also summarized in Table~\ref{tab:results}.

\subsection{Likelihood analysis}
\label{subsec:likelihood}
Following Braun et al.\ \cite{Braun:2008bg} we use a likelihood method to analyze the PSTracks data for point sources. We calculate the probability density $P_j(E_j|\phi^{\rm src})$ for a muon with energy $E_j$ from an astrophysical point source with flux $\phi^{\rm src}$ as   
\begin{eqnarray}
	P(E_j|\phi^{\rm src}) = \frac{\sum_k M(E_j, E^*_k) n_{s,k}}
	{\sum_k n_{s,k}} .
	\label{eq:prob_src_energy}
\end{eqnarray}
Here $n_{s,k}$ is the signal event number calculated using Eq.~(\ref{eq:signal_evts}) for the energy interval $E_k \le E_{\nu} \le E_{k+1}$, and $M(E_j, E^*_k)$ is an energy migration matrix 
to obtain a muon with energy $E_{j}$, from a neutrino in the energy range $E_k \le E_{\nu} \le E_{k+1}$  represented by $E^*_k$,
provided by the IceCube Collaboration as part of the instrument response function \cite{IceCube:2021xar,IceCube:2019cia}. Finally, after introducing Eq.~(\ref{eq:prob_src_energy}) with a Gaussian spatial probability density profile, we can write a source probability density for the $j$-th neutrino event as  
\begin{eqnarray}
	{\cal S}_{j} (\vec{x}_j, \vec{x}_s, E_j, \phi^{\rm src}) 
	= \frac{1}{2\pi\sigma_j^2} e^{-\frac{|\vec{x}_j - \vec{x}_s|^2}{2\sigma_j^2}}
	P(E_j|\phi^{\rm src}).
	\label{eq:prob_src_evt}
\end{eqnarray}
Here $\vec{x}_s$ is a unit vector to the direction of the point source, while $\vec{x}_j$ is that of the $j$-th neutrino arrival direction. These unit vectors can be written in terms of Right Ascension ($\varphi$) and Declination ($\delta$) as 
$$\vec{x} = (\sin\delta \cos\varphi,\, \sin\delta \sin\varphi,\, \cos\delta)^t.$$ 
The term $\sigma_j$ in Eq.~(\ref{eq:prob_src_evt}) is the angular error associated with the $j$-th neutrino event.

Similarly, we calculate the background probability density for the $j$-th $\nu$ event as
\begin{eqnarray}
	{\cal B}_j = \frac{P(E_j|\phi^{\rm atm} + \phi^{\rm ast})}{\Delta\Omega_s},
	\label{eq:prob_bkg_evt}
\end{eqnarray}
where we assume that the background events are uniformly distributed within a solid angle $\Delta\Omega_s$ around the source direction $\vec{x}_s$ in the sky. The solid angle is a square of side $12^\circ$ centered at a particular source, motivated by Ref.~\cite{Braun:2009wp}.  We calculate $P(E_j|\phi^{\rm atm} + \phi^{\rm ast})$ in the same way as equation (\ref{eq:prob_src_energy}) but by replacing the source flux with the background fluxes to calculate the background events $n_{b,k}$. Finally, for a total of $N$ neutrino events within $\Delta\Omega_s$, we compute a likelihood function as
\begin{eqnarray}
	{\cal L} (\vec{x}_s; {\theta}) = \prod_{j = 1}^N \left[ 
	\frac{n_s}{N} {\cal S}_j + \left( 1 - \frac{n_s}{N} \right) {\cal B}_j
	\right],
	\label{eq:likelihood}
\end{eqnarray}
where ${\theta} = \{ n_s, \gamma, \delta m^2\}$ is the set of free parameters that we vary to maximize $\log {\cal L}$. Note that, for the parameter set ${\theta} = \{ n_s, \gamma, \delta m^2 = 0\}$,  the pseudo-Dirac scenario is converted to the conventional 3-$\nu$ oscillation scenario. 
We define a test statistic based on the likelihood ratio test as
\begin{eqnarray}
	{\rm TS} = - 2 \left[ \log{\cal L}(\vec{x}_s; 0)  - \log{\cal L}(\vec{x}_s; {\hat\theta}) \right] ,
	\label{eq:TS}
\end{eqnarray}
where ${\cal L}(\vec{x}_s; 0)$ correspond to the null hypothesis of no signal event, i.e., $n_s=0$ and ${\hat\theta}$ corresponds to the set of parameters for which $\log {\cal L} (\vec{x}_s; {\theta})$ is the maximum. We define a quantity
\begin{equation}\label{eq:DeltaTSprime}
    \Delta {\rm TS}^\prime (n_s, \gamma) = {\rm TS} (\hat{n}_s, \hat{\gamma}) - {\rm TS} (n_s, \gamma)
\end{equation}
to obtain the best fit values of $n_s$ and $\gamma$ in both SM and pseudo-Dirac (pD) scenarios.
To distinguish between the SM and pseudo-Dirac scenarios and to constrain $\delta m^2$ we use the difference between the TS values for the two cases as 
\begin{eqnarray}
	{\Delta} {\rm TS} = {\rm TS}^{\rm SM} - {\rm TS}^{\rm pD} , 
	\label{eq:DTS}
\end{eqnarray}
where ${\rm TS}^{\rm SM}$ correspond to $\delta m^2 = 0$.  ${\Delta} {\rm TS}$ is expected to be distributed as a $\chi^2$ with one degree of freedom.

\section{Results}
\label{sec:results}

\begin{figure*}[htbp]
	\centering
	\includegraphics[width=.34\textwidth]{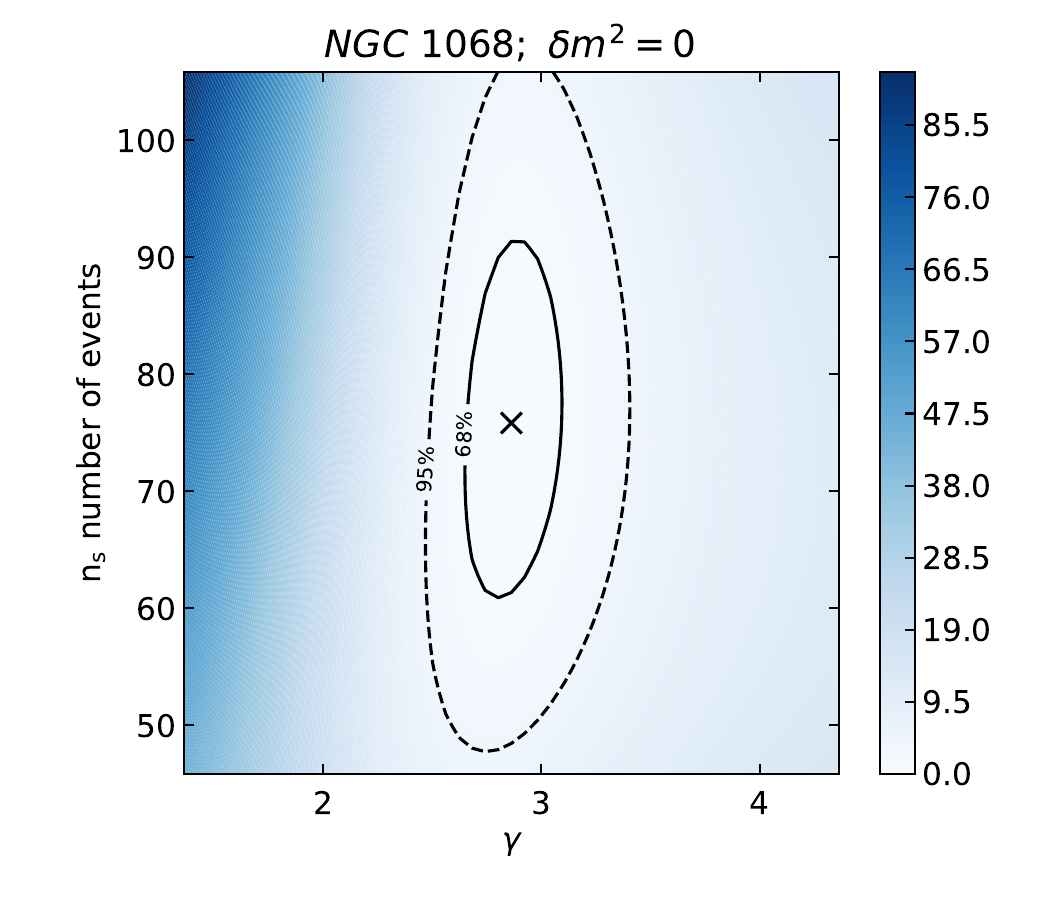}
	\includegraphics[width=.34\textwidth]{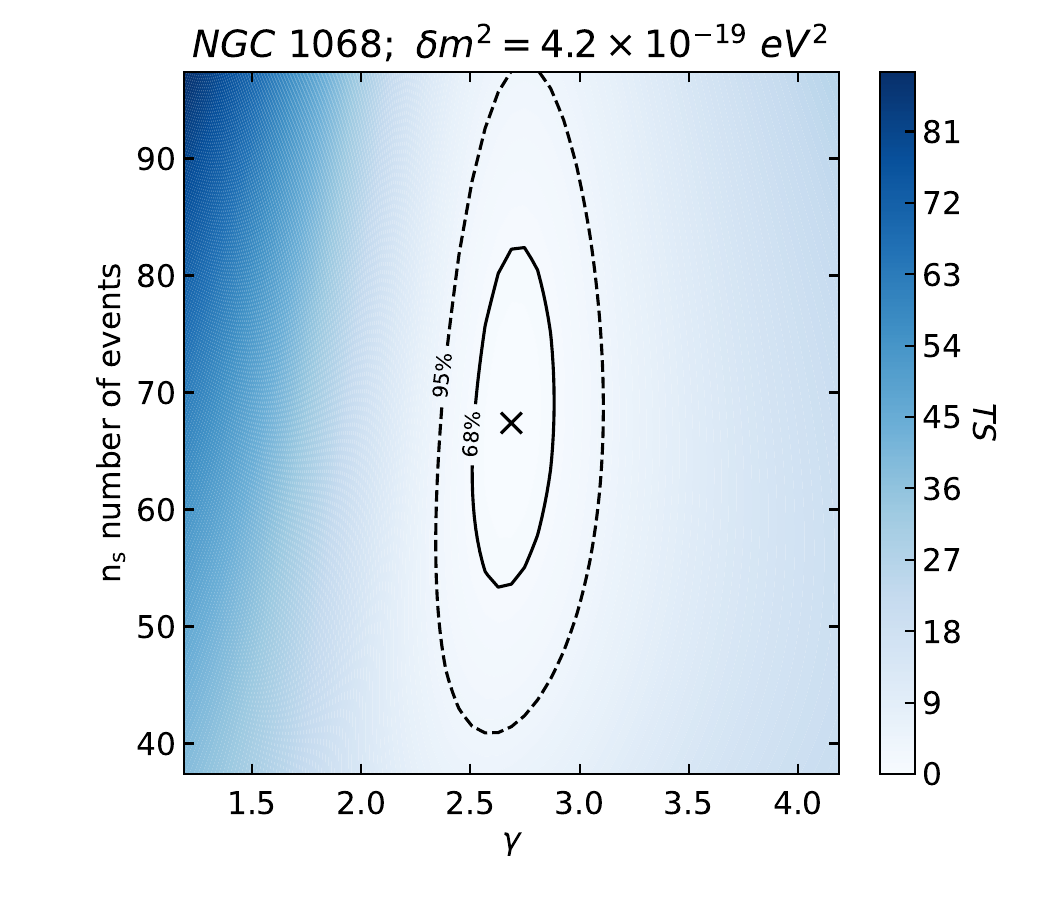}\\
	\includegraphics[width=.34\textwidth]{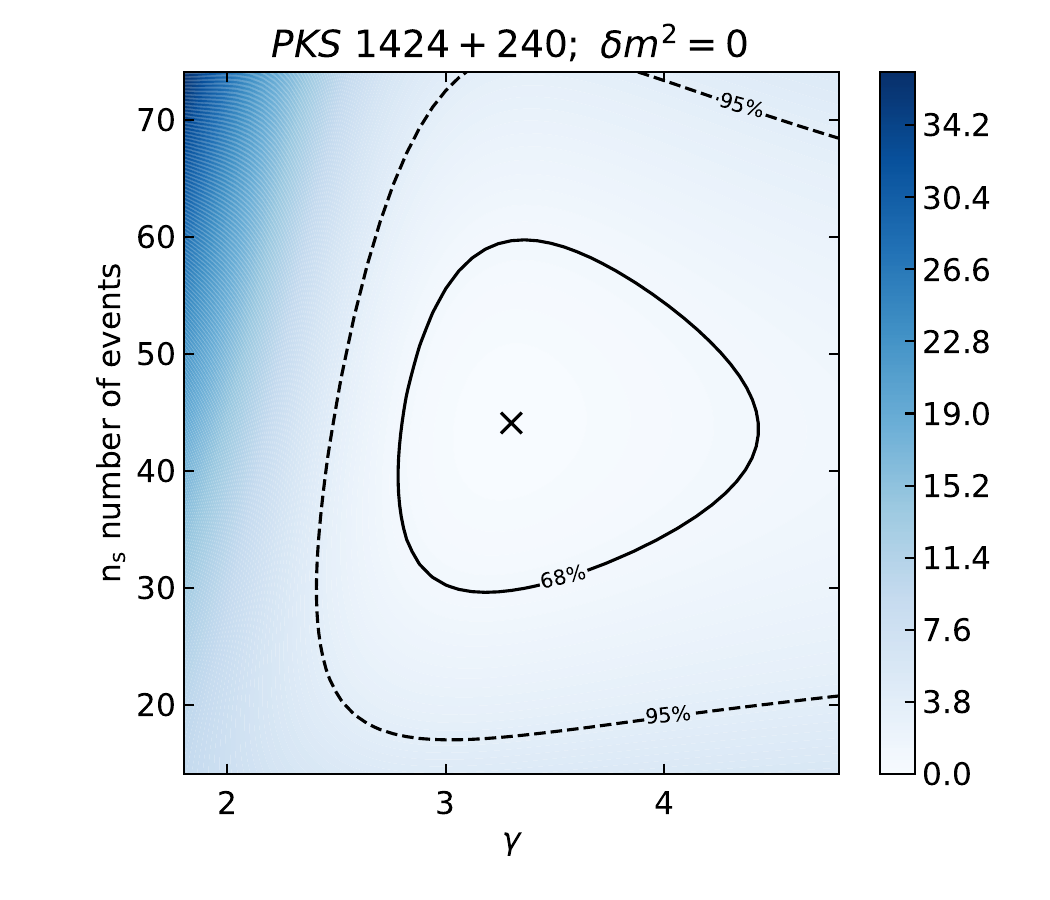}
	\includegraphics[width=.34\textwidth]{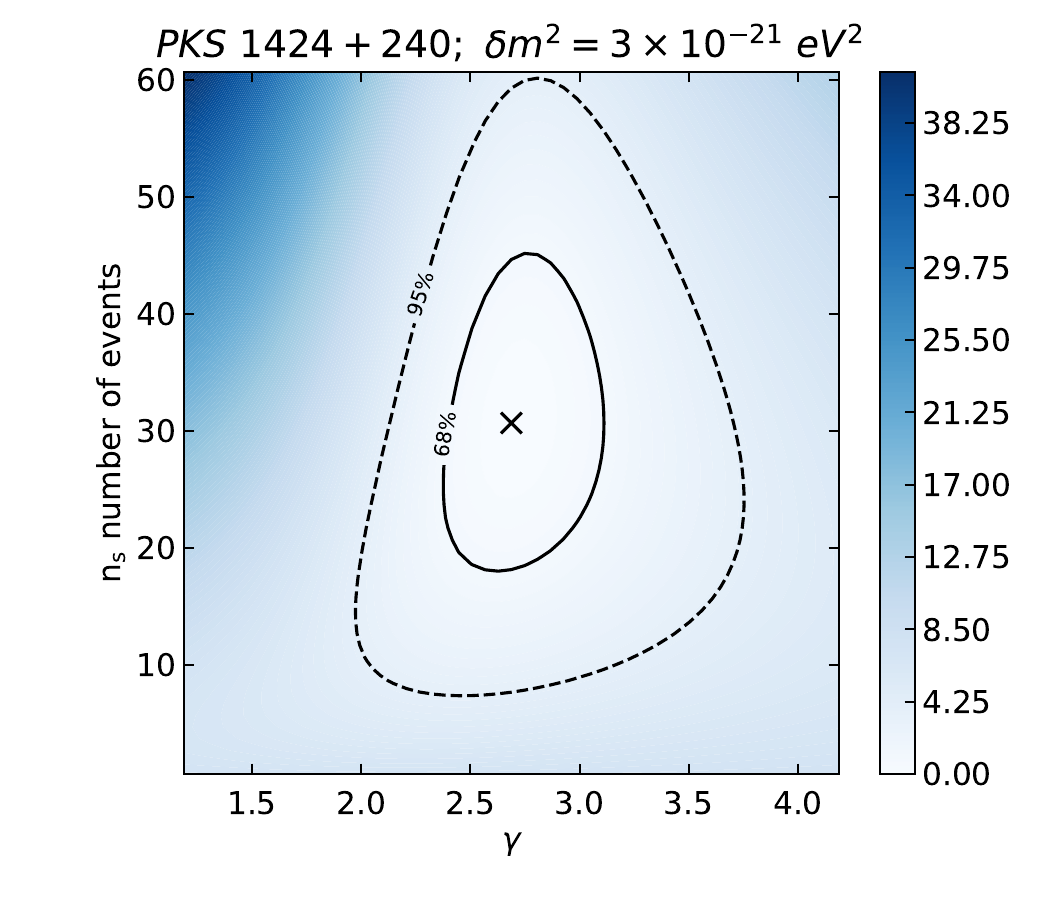}\\
	\includegraphics[width=.34\textwidth]{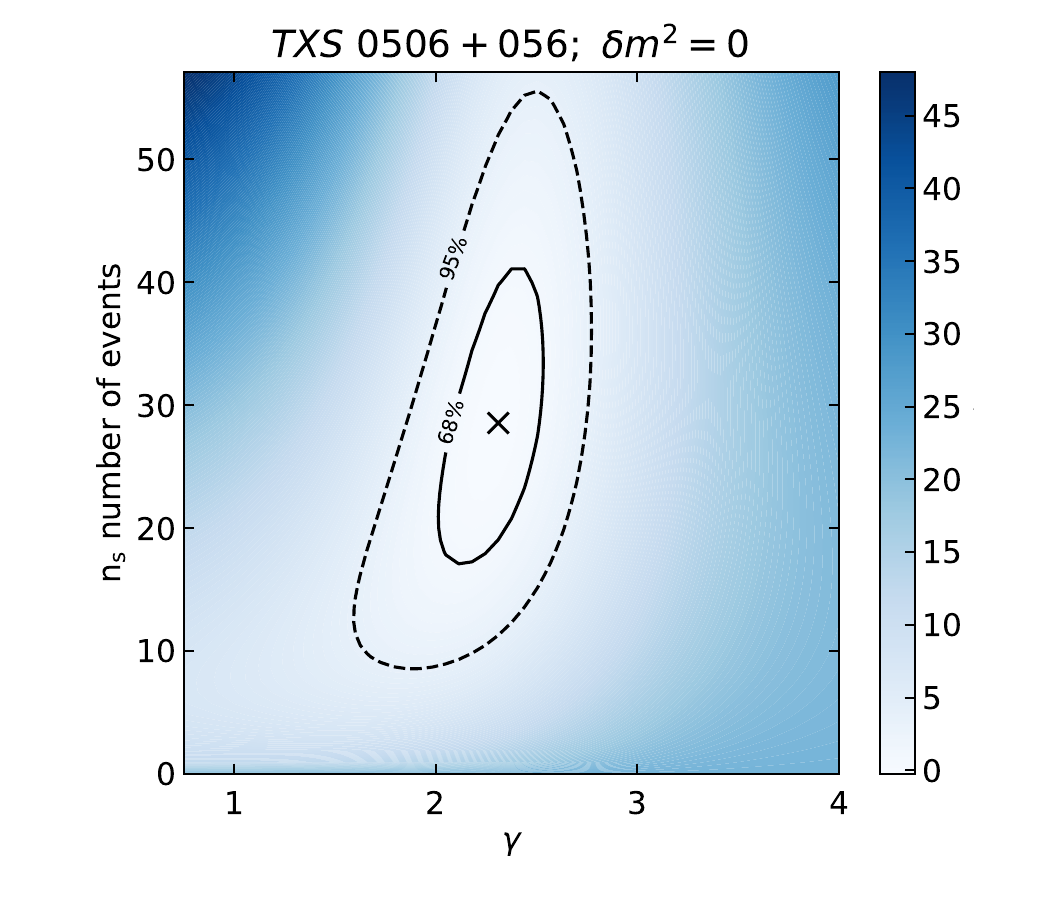}
	\includegraphics[width=.34\textwidth]{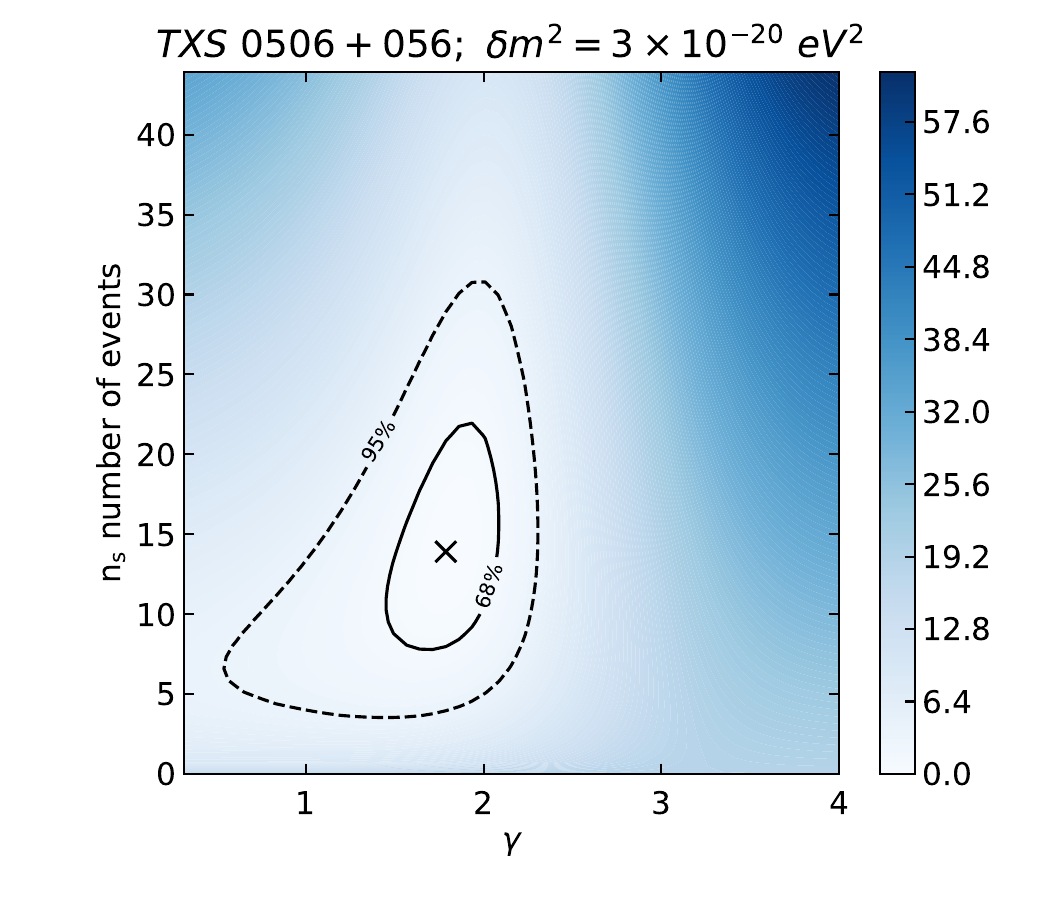}\\
	\includegraphics[width=.34\textwidth]{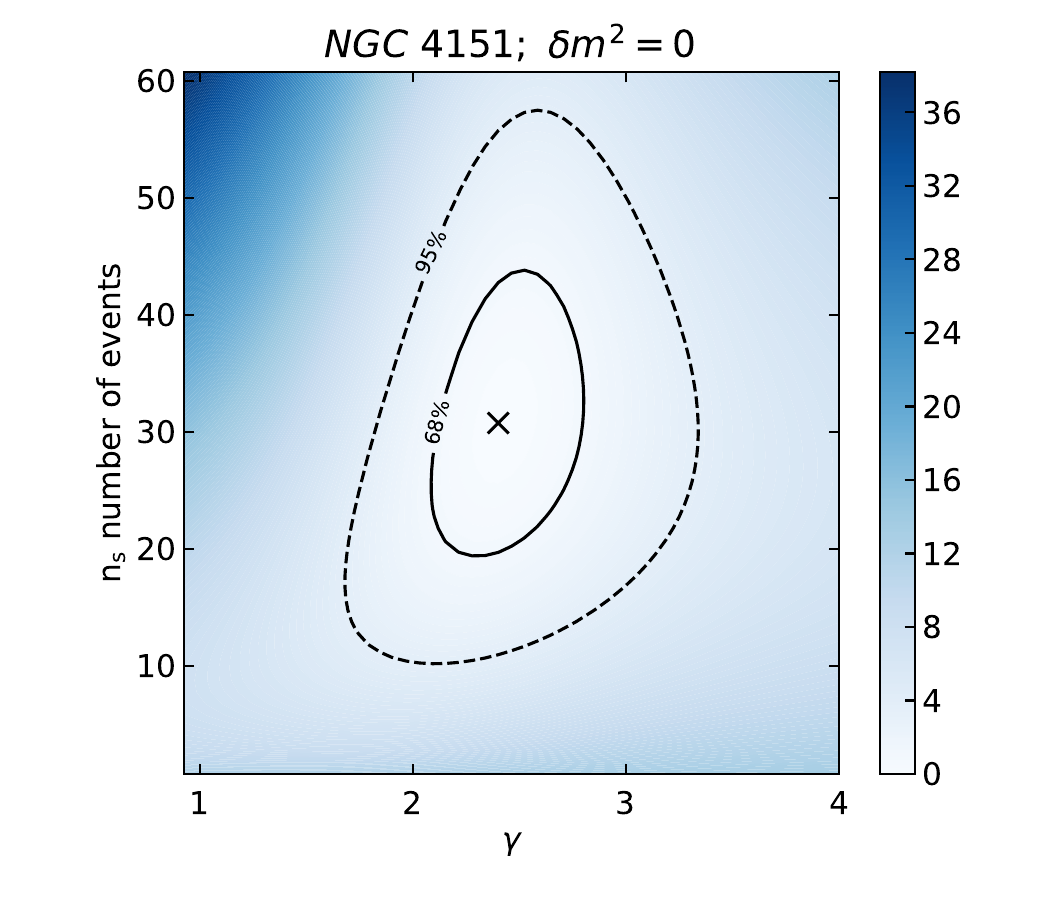}
	\includegraphics[width=.34\textwidth]{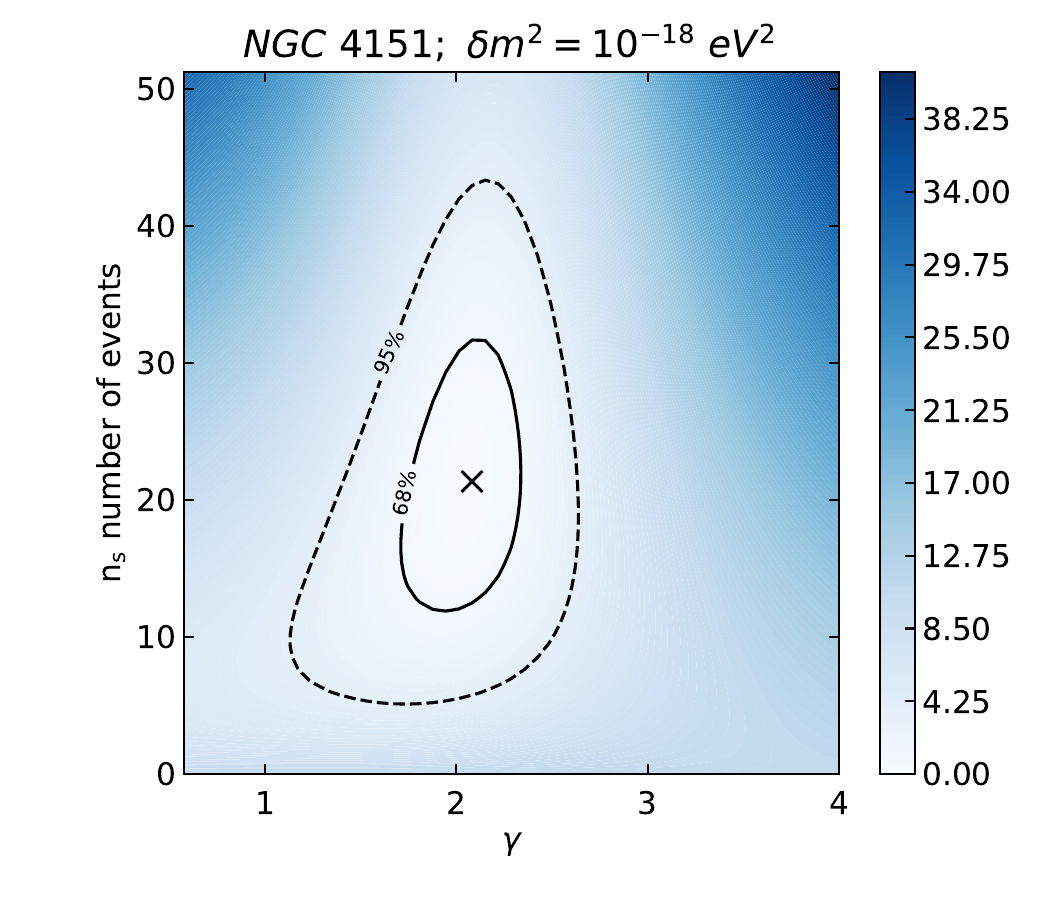}
	\caption{Contour plots of the TS scanned over the parameter $n_s$ and $\gamma$ and subtracted from the maximum TS obtained for their best-fit values ${\hat n}_s$ and $\hat\gamma$ (the quantity $\Delta {\rm TS}^\prime$ defined in Eq. (\ref{eq:DeltaTSprime}) represented by the color bar), have been projected in the $\gamma$-$n_s$ plane for both $\delta m^2 = 0$ (SM) and for a nonzero $\delta m^2$ value for the sources (except GB6 J1542+6129) listed in Table \ref{tab:sources}. The best-fit point in each case is represented with `x' and the allowed regions with 68\% and 95\% C.L. are shown as solid and dashed curves, respectively. The energy-range for IceCube is considered to be 0.5 TeV - 1 PeV.}
	\label{fig:contours} 
\end{figure*}

In this section, we discuss our results based on the approach defined for the statistical analysis in the previous section. 
In Fig.~\ref{fig:contours}, $\Delta {\rm TS}^\prime$ defined in Eq.~(\ref{eq:DeltaTSprime}) 
is projected in the ($\gamma$ - $n_s$)-plane both in the case of standard (left panels) and pseudo-Dirac neutrinos (right panels) in the context of four sources with the best-fit points (black `x') and the corresponding 68\% and 95\% significance regions. We chose specific values of $\delta m^2$ for each source that fall within the range of (0.1 - 0.5)$L_{\rm osc}$, where $L_{\rm osc}$ represents the oscillation length associated with active-sterile oscillations. These chosen values of $\delta m^2$ are for illustration purpose. It can be seen that the presence of non-zero $\delta m^2$ parameter significantly affects the best-fit values of the spectral index as well as the number of events. In general the number of signal events $n_s$ decreases in the pseudo-Dirac scenario as active neutrinos are converted into sterile neutrinos.

In the standard scenario, the best-fit parameter values (${\hat n}_s$, $\hat{\gamma}_{\rm SM}$), as well as the derived value for flux normalization ${\phi}_0$, with their errors and the maximum TS values are listed in Table~\ref{tab:best_fit_values}. Our best fit values of $n_s$ and the corresponding $\phi_0$ are aligned with the observations found in \cite{IceCube:2019cia} that used the data collected during 2008-2018. For NGC 1068 and NGC 4151 sources, the $n_s$ values also match with the values provided in \cite{IceCube:2022der} and \cite{IceCube:2024ayt}, respectively. Any small differences are likely due to different threshold values used in our paper compared to \cite{IceCube:2019cia} and due to the fact that we are using the public data release. 
The $n_s$ values are shifted to ${\hat n}_s \approx 67$, 30, 14 and 21; and ${\hat \gamma} \approx 2.7$, 2.7, 1.7 and 2.1 for NGC 1068, PKS 1424+240, TXS 0506+056 and NGC~4151 in the pseudo-Dirac scenario for the particular $\delta m^2$ values mentioned in the Fig \ref{fig:contours}. 
We can notice here that the spectral index shifts to lower values, $i.e.,$ the spectrum gets harder for non-zero value of $\delta m^2$. This can be attributed to the fact that in the pseudo-Dirac scenario, fast oscillations at lower energies are averaged out to a lower value compared to the standard oscillations. This results in a change in the flux reaching the earth, see e.g., Fig.~1 in Ref.~\cite{Rink:2022nvw}, which is, in general, lower in normalization and harder in the index compared to the standard case.  The event distribution changes are discussed in more detail later on and relevant plots are shown in the appendix.

\begin{figure*}[htbp]
\centering	\includegraphics[width=.45\textwidth]{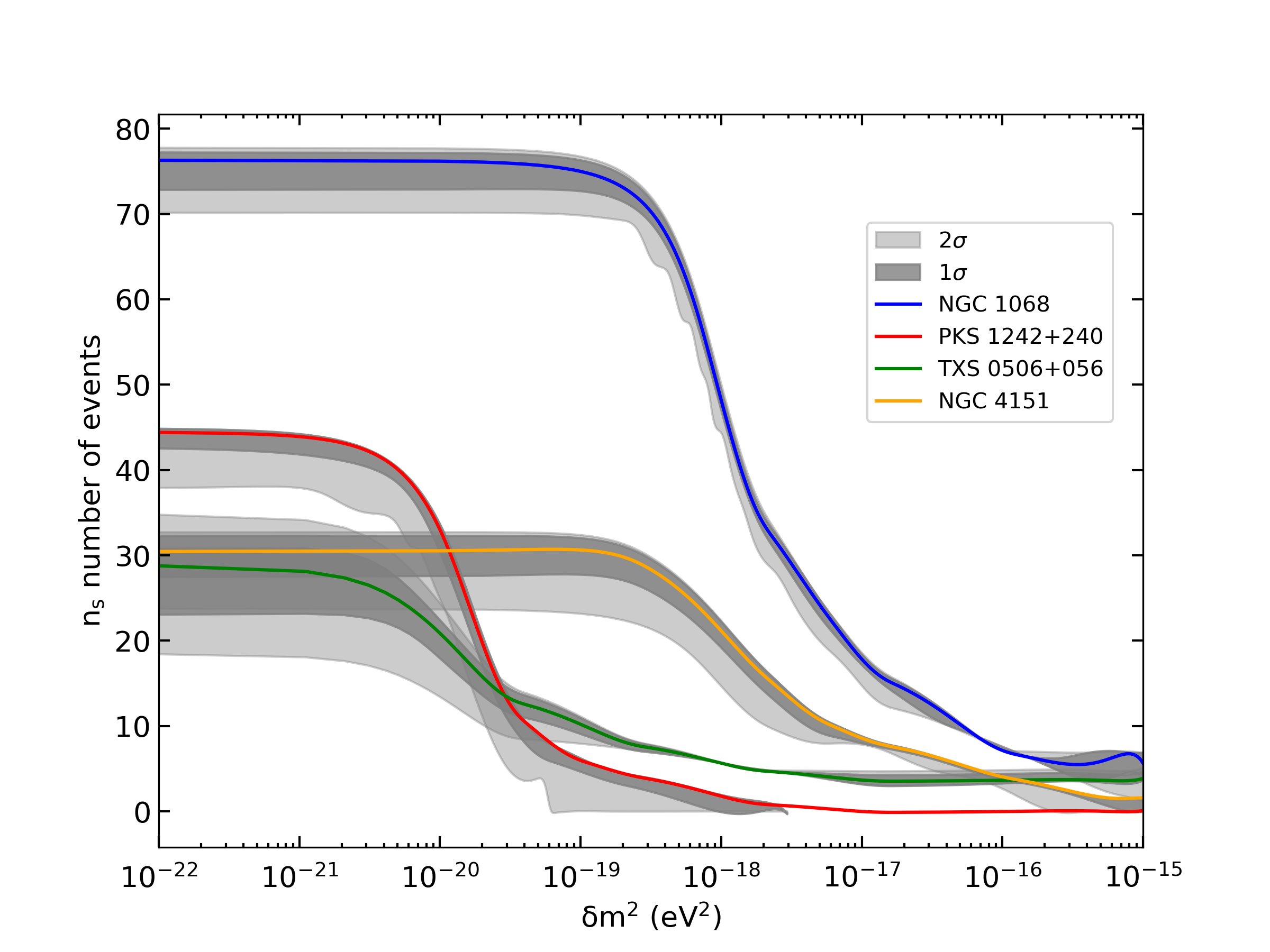}
\includegraphics[width=.45\textwidth]{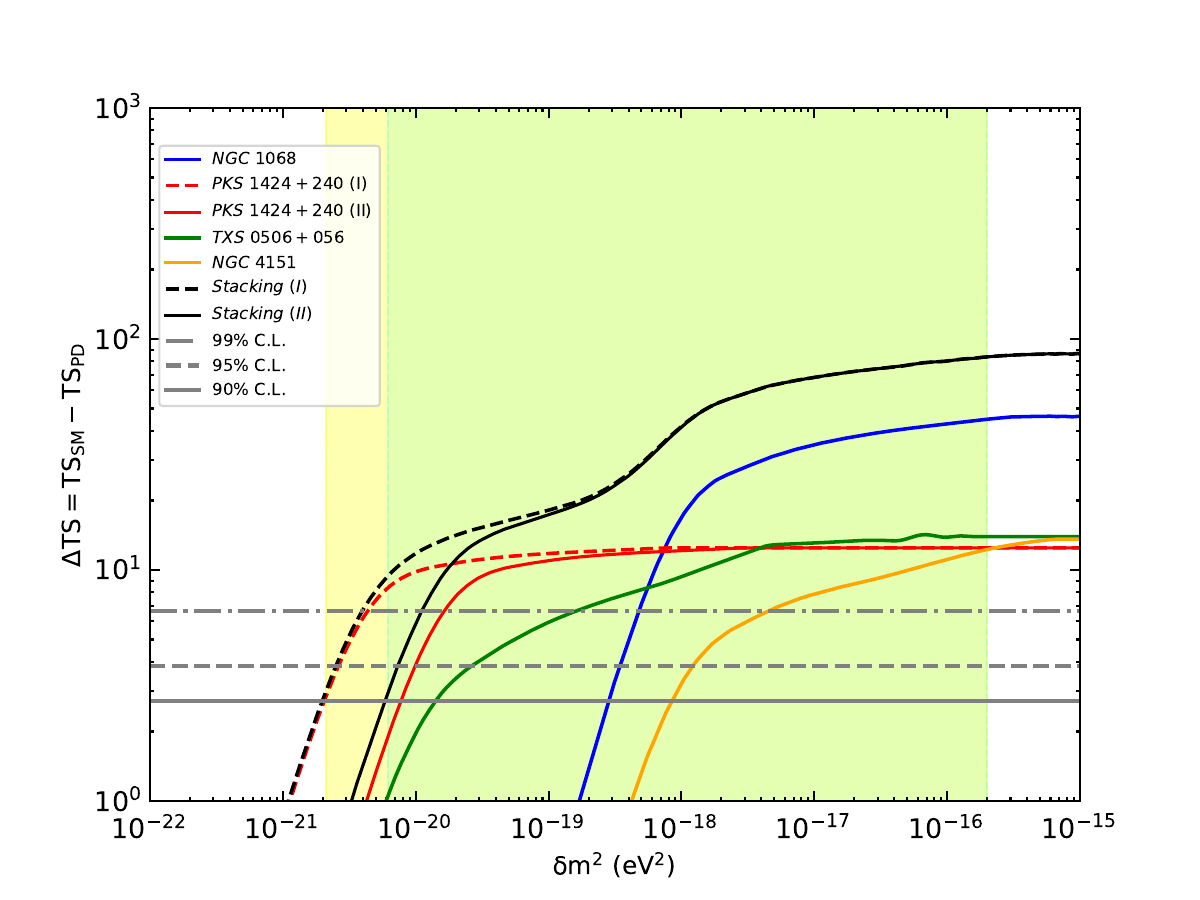}
	\caption{The results of the statistical analysis are presented in this figure. In the left panel, the number of events with respect to $\delta m^2$ are shown for all four sources. The best-fit curves are shown in solid blue (NGC 1068), red (PKS 1424+240), green (TXS 0506+056) and orange (NGC~4151), along with their 1$\sigma$ and 2$\sigma$ regions in dark and light gray shades, respectively. In the right panel, the curves representing $\Delta$TS vs. $\delta m^2$ are shown. We have provided the constraints on $\delta m^2$ parameter at 90\%, 95\% and 99\% C.L.\ from different sources individually with blue-solid (NGC 1068), red-dashed (PKS 1424+240 I), red-solid (PKS 1424+240 II), green-solid (TXS 0506+056) and orange-solid (NGC~4151) curves. Results from the stacking analysis are represented as black-dashed (includes PKS 1424+240 I) and black-solid (includes PKS 1424+240 II) curves, for the two redshift values of PKS 1424+240. The yellow and green shaded regions corresponding to the stacking analysis I and II, respectively, represent the $\delta m^2$ range excluded at $\ge 90\%$ CL. For this analysis, we considered the neutrino energy range as 0.5 TeV - 1 PeV.}
	\label{fig:stacking} 
\end{figure*}

In the left panel of Fig.~\ref{fig:stacking}, variation of the number of events is shown as a function of $\delta m^2$ for the four sources considered. The solid blue, red, green and orange curves exhibit the best-fit values for NGC 1068, PKS 1424+240, TXS 0506+056 and NGC~4151 with corresponding 1$\sigma$ and 2$\sigma$ regions in darker and lighter shades, respectively, of gray color.

Before we put constraints on $\delta m^2$, we note that there is also a threshold above which the detectors become sensitive to the difference between the standard and pseudo-Dirac oscillations. These threshold $\delta m^2$ values ($\delta m^2_{\rm th}$) can be roughly estimated from the probability curves such as those plotted in Fig.~\ref{fig:probabilities_dmsqr}, showing 1\% change between the standard and pseudo-Dirac probabilities, for different energies and are listed in Table~\ref{tab:results}.

To constrain the $\delta m^2$ parameter, we obtained Fig.~\ref{fig:stacking} right panel using $\Delta$TS defined in Eq.~(\ref{eq:DTS}). 
The blue, red, green and orange solid curves represent the case of NGC 1068, PKS 1424+240, TXS 0506+056 and NGC~4151, respectively. The black curve represents the stacking analysis combining the data from all these four sources. The 90\%, 95\% and 99\% C.L.\ are shown as solid, dashed and dot-dashed curves, respectively, in gray color. For PKS 1424+240, there are two possible redshift values, $z = 0.604$ and $z = 0.16$. Hence, we incorporated the analysis for this source for both the redshifts. The results are shown in red dashed (for $z = 0.604$) and red solid (for $z = 0.16$) curves, we call these analysis-I and analysis-II, respectively. Moreover, the stacking analysis is also shown as black dashed (analysis-I) and black solid (analysis-II) curves for the given two values of the redshift for PKS 1424+240. During this analysis, we also kept the number of signal events $n_s$ and spectral index $\gamma$ as free parameters. 
The constraints we have obtained on $\delta m^2$ for different sources and from stacking analyses are reported in Table~\ref{tab:results} for both the 0.5 TeV - 1 PeV and 0.1 TeV - 1 PeV energy ranges. 

Note, however, that these constraints apply only in a limited $\delta m^2$ range as after a certain upper value of $\delta m^2$ the likelihood ratio no longer changes with $\delta m^2$. It stems from the fact that for large $\delta m^2$ values the rapid oscillations between active and sterile neutrinos get averaged out. The analysis still retains sensitivity to the overall normalization, but the oscillatory modulation is averaged out, which restricts the analysis to examining a specific range of $\delta m^2$.
To fix this upper limit we require that $\Delta$TS does not increase by more than 1\% for $\delta m^2 > \delta m^2_{\rm max}$, see Ref.~\cite{Rink:2022nvw}. 
Therefore, our exclusion region is between a minimum value $\delta m_{\rm min}^2$ when $\Delta$TS crosses the 90\% CL limit and up to $\delta m^2_{\rm max}$.

The plateau region appears at different $\delta m^2$ values for different sources, PKS~1424+240 provides the lowest values for both $\delta m^2_{\rm min}$ and $\delta m^2_{\rm max}$, and the narrowest range of $\delta m^2$ excluded at $\ge 90\%$ CL. The stacking analysis, however, provides a higher value of $\delta m^2_{\rm max}$, dominated by NGC~1068, and a broader range of exclusion. 
The exclusion ranges are reported in table~\ref{tab:results} for each source, 
and also shown in the right panel of Fig.~\ref{fig:stacking} as shaded regions. Conclusively, the excluded regions after stacking analysis are $\delta m^2 \in [2.1\times 10^{-21} - 2\times 10^{-16}]$  eV$^2$ for stacking analysis (I) and $\delta m^2 \in [6.2\times 10^{-21} - 2\times 10^{-16}]$ eV$^2$ for stacking analysis (II), for the neutrino energy range of 0.5 TeV - 1 PeV.

In Figs.~\ref{fig:event_0pt5to1000TeV} and \ref{fig:event}, given in the appendix, event distribution is provided for all four sources for SM scenario ($\delta m^2 = 0$, black-solid line) and for the pseudo-Dirac scenario with two different non-zero values of $\delta m^2$ corresponding to 90\% and 95\% C.L. as red and blue curves, respectively. 
There is a change in the event distribution for $\delta m^2 \neq 0$, as the oscillation in the pseudo-Dirac scenario changes the flux reaching the earth. This change affects both the normalization and spectral shape, in general making the index harder as discussed previously. As a result, the event distribution changes, resulting in fewer events at lower energies and more events at higher energies compared to the standard oscillation scenario.

The results discussed so far are regarding the 0.5 TeV - 1 PeV range considered for IceCube data, while we have also performed the same analysis for 0.1 TeV - 1 PeV range. The results are summarized in Table \ref{tab:results}, also the event distribution for both energy ranges are given in the appendix. Contour plots as well as the $\Delta$TS vs. $\delta m^2$ and $n_s$ (number of events) vs. $\delta m^2$ for 0.1 TeV - 1 PeV are also provided in the appendix as Fig.~\ref{fig:contours_0pt1to1000TeV} and Fig.~\ref{fig:stacking2}, respectively. It can be seen that the lower limit of the excluded range is deceased by lowering the threshold energy of neutrinos. The exclusion window in this case is $\delta m^2 \in [1.1 \times 10^{-21} - 3 \times 10^{-16}]$ eV$^2$ and $\delta m^2 \in [3 \times 10^{-21} - 3 \times 10^{-16}]$ eV$^2$, from stacking analysis (I) and (II), respectively.

\begin{table*}[htbp]
	\centering
	\begin{tabular}{|l|c|c|c|c|c|c|c|c|}
		\hline\noalign{\smallskip}
		\makecell{Energy\\range} & & \multicolumn{1}{c|}{NGC 1068} & \multicolumn{1}{c|}{\makecell{TXS\\ 0506+056}} & \multicolumn{1}{c|}{\makecell{PKS\\ 1424+240\\ (I)}} & \multicolumn{1}{c|}{\makecell{PKS\\ 1424+240\\ (II)}} & \multicolumn{1}{c|}{NGC 4151} & \multicolumn{1}{c|}{\makecell{Stacking\\ (I)}} & \multicolumn{1}{c|}{\makecell{Stacking\\ (II)}} \\
		\hline\noalign{\smallskip}
		& \makecell{$\delta m^2_{\rm th}{\rm (eV^2)}$} & $9 \times 10^{-20}$ & $9 \times 10^{-22}$ &$5 \times 10^{-22}$ & $2 \times 10^{-21}$ & $9.1 \times 10^{-20}$ & $5 \times 10^{-22}$ & $2 \times 10^{-21}$ \\
		\makecell{0.5 TeV\\ - 1 PeV\\ (Fixed\\ Back-\\ ground)} & \makecell{$\delta m^2_{\rm min}{\rm (eV^2)}$} & $2.9 \times 10^{-19}$ & $1.5 \times 10^{-20}$ &$2.1 \times 10^{-21}$ & $7.5 \times 10^{-21}$ & $8.5 \times 10^{-19}$ & $2.1 \times 10^{-21}$ & $6.2 \times 10^{-21}$ \\
        & $\delta m^2_{\rm max}{\rm (eV^2)}$  & $2.0 \times 10^{-16}$ & $1.0 \times 10^{-17}$ & $1.0 \times 10^{-18}$ & $2.0 \times 10^{-18}$ & $8.0 \times 10^{-16}$ & $2.0 \times 10^{-16}$ & $2.0 \times 10^{-16}$ \\
		\addlinespace
		\hline
		\addlinespace
		& \makecell{$\delta m^2_{\rm th}{\rm (eV^2)}$} & $2 \times 10^{-20}$ & $2 \times 10^{-22}$ &$1 \times 10^{-22}$ & $4 \times 10^{-22}$ & $2.1 \times 10^{-20}$ & $1 \times 10^{-22}$ & $4 \times 10^{-22}$ \\
		\makecell{0.1 TeV\\ - 1 PeV\\ (Fixed\\ Back-\\ ground)} & \makecell{$\delta m^2_{\rm min}{\rm (eV^2)}$} & $1.3 \times 10^{-19}$ & $1.4 \times 10^{-20}$ &$1.1 \times 10^{-21}$ & $3.9 \times 10^{-21}$ & $6.5 \times 10^{-19}$ & $1.1 \times 10^{-21}$ & $3.0 \times 10^{-21}$ \\
        &  $\delta m^2_{\rm max}{\rm (eV^2)}$  & $3 \times 10^{-16}$ & $1.0 \times 10^{-17}$ & $1.0 \times 10^{-18}$ & $3.0 \times 10^{-18}$ & $8.0 \times 10^{-16}$ & $3.0 \times 10^{-16}$ & $3.0 \times 10^{-16}$ \\
		\addlinespace
		\hline
		\addlinespace
		\makecell{\\0.5 TeV\\ - 1 PeV\\ (Data\\-driven\\ Back-\\ ground)} & \makecell{$\delta m^2_{\rm min}{\rm (eV^2)}$} & $2.3 \times 10^{-19}$ & $1.1 \times 10^{-19}$ & $1.5 \times 10^{-21}$ & $5.9\times 10^{-21}$ & $1.2 \times 10^{-18}$ & $1.5 \times 10^{-21}$ & $5.5 \times 10^{-21}$ \\
        & $\delta m^2_{\rm max}{\rm (eV^2)}$  & $3 \times 10^{-16}$ & $1.0 \times 10^{-17}$ & $2.0 \times 10^{-20}$ & $1.0 \times 10^{-19}$ & $8 \times 10^{-16}$ & $3.0 \times 10^{-16}$ & $3.0 \times 10^{-16}$ \\
		\addlinespace
		\hline
	\end{tabular}
	\caption{Results of the analysis performed for all sources and the stacking analysis are given in terms of constraints of $\delta m^2$ with $\ge 90\%$ CL within the range $\delta m_{\rm min}^2 - \delta m_{\rm max}^2$. Also shown in the table are the $\delta m^2$ values ($\delta m^2_{\rm th}$) above which the sensitivity to the pseudo-Dirac neutrinos can be probed.}
	\label{tab:results}
\end{table*}

\begin{table*}[htbp]
\centering
\small 
\begin{tabular}{llcccc}
\toprule
  & Source & $\hat{\gamma}_{\rm SM} \pm 1\sigma$ & ${\hat n}_s \pm 1\sigma$ & ${\hat \phi}_0 \pm 1\sigma ~({\rm eV}^2)$ & TS$_{\rm max}$ \\
\midrule
\multirow{4}{*}{\makecell{Fixed\\ background \\ (0.5~TeV - 1~PeV)}} 
& NGC~1068 & $2.9^{+0.2}_{-0.2}$ & $76^{+16}_{-15}$ & $4.42^{+0.95}_{-1.11}\times 10^{-11}$ & 48.39 \\
\addlinespace
& TXS~0506+056 & $2.3^{+0.2}_{-0.3}$ & $28^{+14}_{-12}$ & $0.74^{+0.69}_{-0.49}\times 10^{-11}$ & 22.78 \\
\addlinespace
& PKS~1424+240 & $3.3^{+1.2}_{-0.5}$ & $44^{+16}_{-14}$ & $2.75^{+1.0}_{-0.86}\times 10^{-11}$ & 12.45 \\
\addlinespace
& NGC 4151 & $2.3^{+0.5}_{-0.2}$ & $31^{+13}_{-12}$ & $0.89^{+0.6}_{-0.44}\times 10^{-11}$ & 14.04 \\
\midrule
\multirow{4}{*}{\makecell{Fixed\\ background \\ (0.1~TeV - 1~PeV)}} 
& NGC~1068 & $2.75^{+0.2}_{-0.2}$ & $75^{+16}_{-15}$ & $3.98^{+1.52}_{-1.46}\times 10^{-11}$ & 48.39 \\
\addlinespace
& TXS~0506+056 & $2.2^{+0.3}_{-0.2}$ & $28^{+12}_{-12}$ & $0.58^{+0.98}_{-0.4}\times 10^{-11}$ & 22.78 \\
\addlinespace
& PKS~1424+240 & $3.0^{+0.5}_{-0.4}$ & $42^{+16}_{-14}$ & $2.58^{+0.81}_{-1.32}\times 10^{-11}$ & 12.45 \\
\addlinespace
& NGC 4151 & $2.4^{+0.4}_{-0.3}$ & $30^{+13}_{-10}$ & $1.02^{+0.65}_{-0.47}\times 10^{-11}$ & 14.04 \\
\midrule
\multirow{4}{*}{\makecell{Data-driven\\ background \\ (0.5~TeV - 1~PeV)}} 
& NGC~1068 & $3.3^{+0.5}_{-0.3}$ & $75^{+15}_{-15}$ & $4.69^{+0.94}_{-0.9}\times 10^{-11}$ & 40.68 \\
\addlinespace
& TXS~0506+056 & $2.2^{+0.3}_{-0.5}$ & $18^{+12}_{-10}$ & $0.37^{+0.42}_{-0.33}\times 10^{-11}$ & 16.32 \\
\addlinespace
& PKS~1424+240 & $4.2^{+1.2}_{-1.0}$ & $42^{+15}_{-13}$ & $1.41^{+0.7}_{-0.06}\times 10^{-11}$ & 11.30 \\
\addlinespace
& NGC 4151 & $2.5^{+0.6}_{-0.5}$ & $23^{+12}_{-11}$ & $0.89^{+0.62}_{-0.61}\times 10^{-11}$ & 8.30 \\
\bottomrule
\end{tabular}
\caption{
Best-fit values with corresponding 1$\sigma$ intervals of $\hat{\gamma}_{\rm SM}$, $\hat{n}_s$ and the derived flux normalization $\phi_0$ in the cases of both fixed and data-driven background in the standard oscillation scenario.}
\label{tab:best_fit_values}
\end{table*}

\subsection{Analysis including data-driven background}

So far, our analysis used fixed background as coming from the atmospheric and diffuse astrophysical neutrinos by employing theoretical models. However another possibility is to use the data driven background analysis. Under this idea, it is considered that the signal is a minor contribution to the total data and the detected events are used as the background, see, e.g., Refs.~\cite{IceCube:2010nca}, \cite{IceCube:2013kvf}, \cite{IceCube:2016tpw}. 
These events are assumed to be uniformly distributed per unit solid angle within $\Delta\Omega_s$ as before in Eq.~(\ref{eq:prob_bkg_evt}). The background probability distribution is then written as
\begin{eqnarray}
	{\cal B}_j =  \frac{N_{j}/N}{{\Delta\Omega_s}},
	\label{eq:prob_bkg_evt2}
\end{eqnarray}
where $N_{j}$ and $N$ are the number of events inside an energy bin where $E_{j}$ belongs, and the total number of neutrinos; inside the square of side 12 degrees centered on the studied source position, respectively. $\Delta\Omega_s$ is the element of solid angle. We found that the background is typically a few orders of magnitude higher than the signal in our conservative $12^\circ \times 12^\circ$ search region, thus validating use of this method.

\begin{figure*}[htbp]
	\centering
	\includegraphics[width=.37\textwidth]{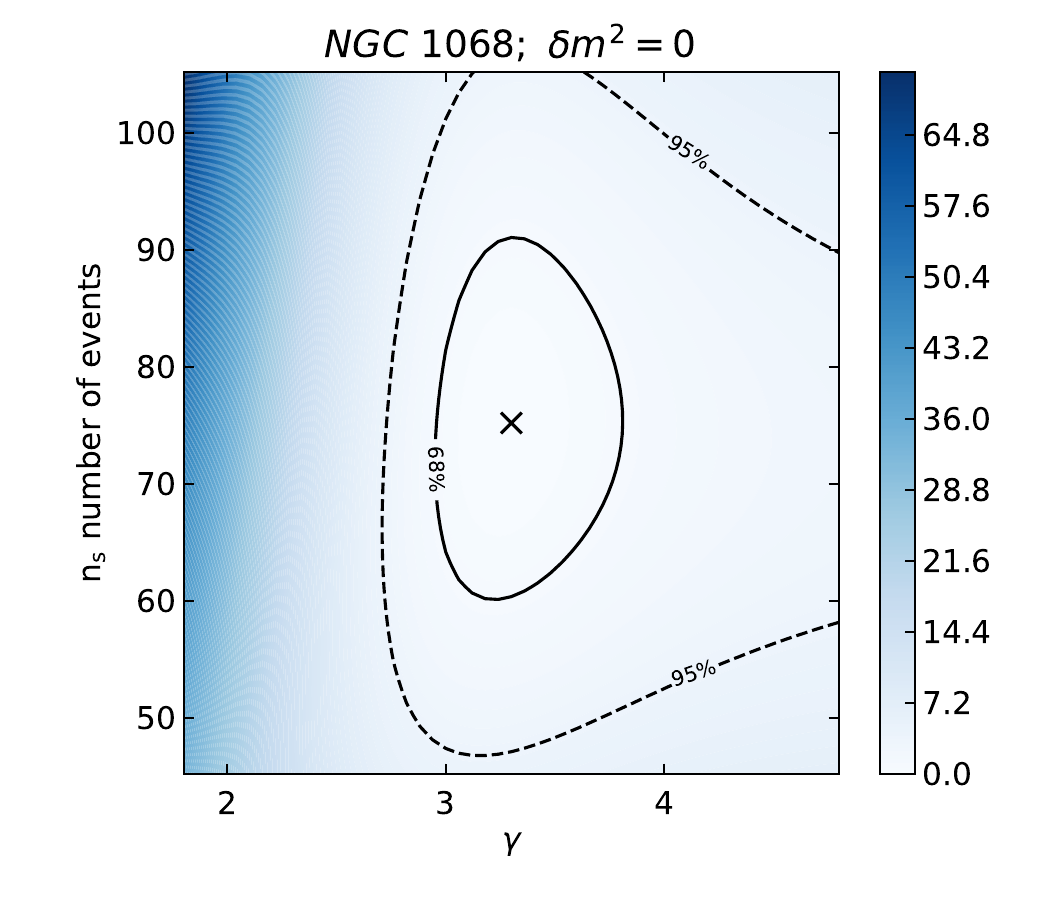}
	\includegraphics[width=.37\textwidth]{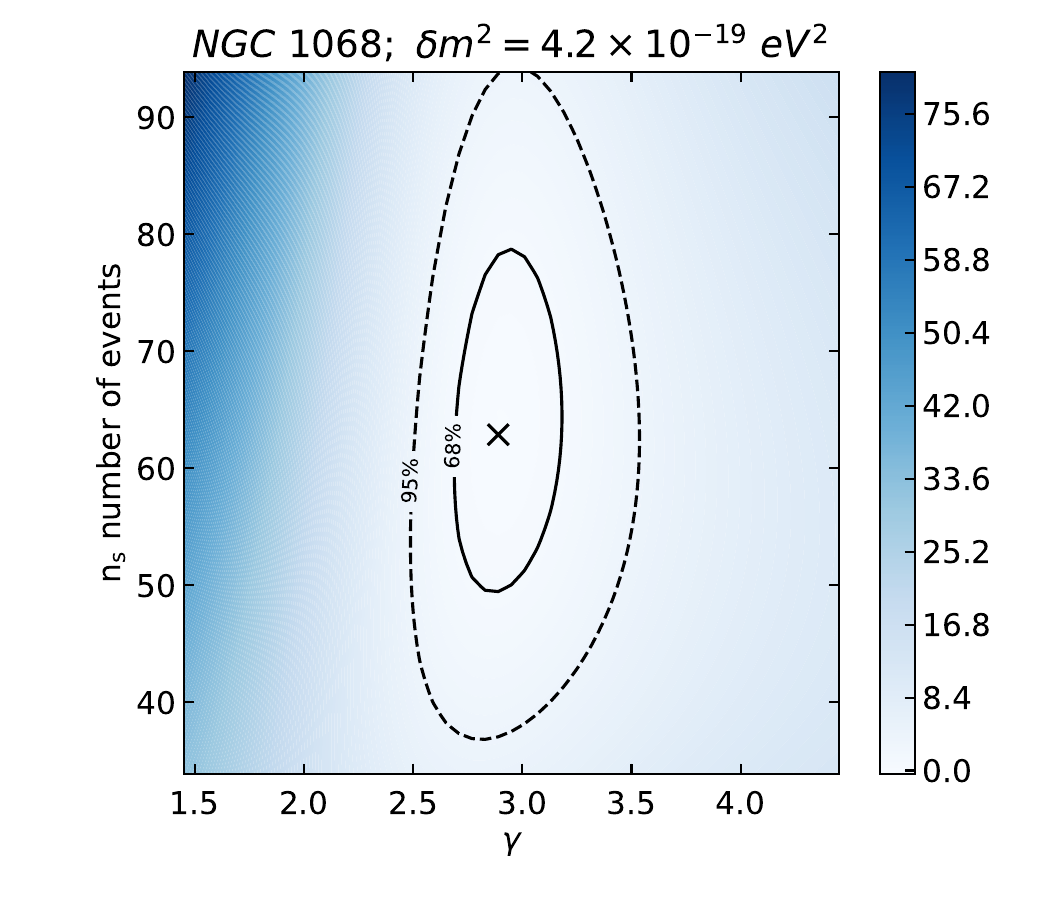}\\
	\includegraphics[width=.37\textwidth]{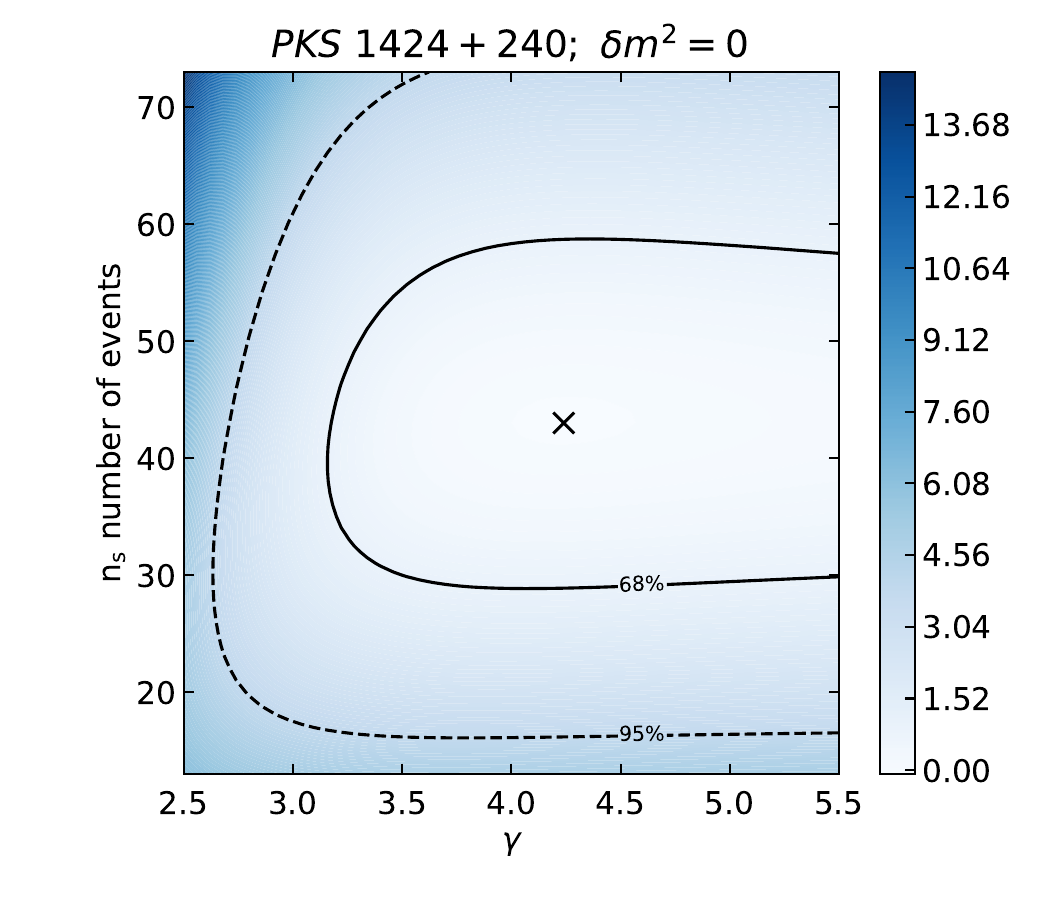}
	\includegraphics[width=.37\textwidth]{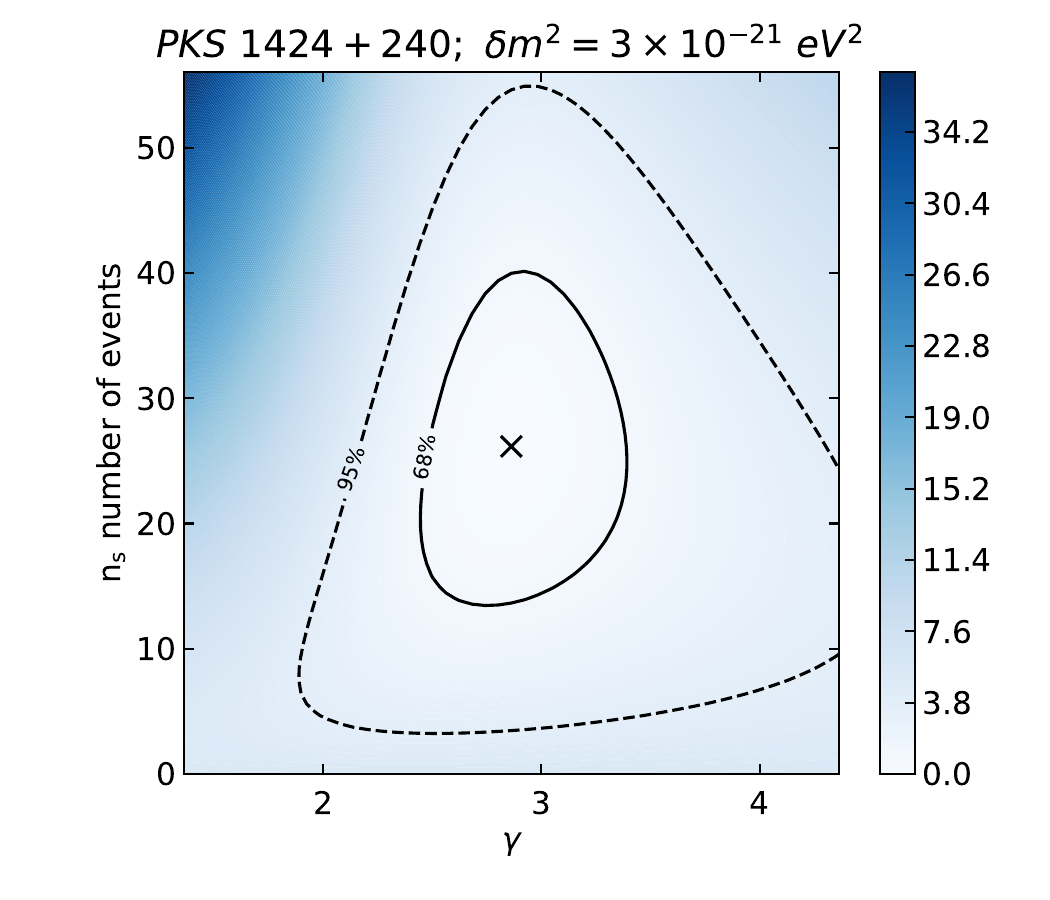}\\
	\includegraphics[width=.37\textwidth]{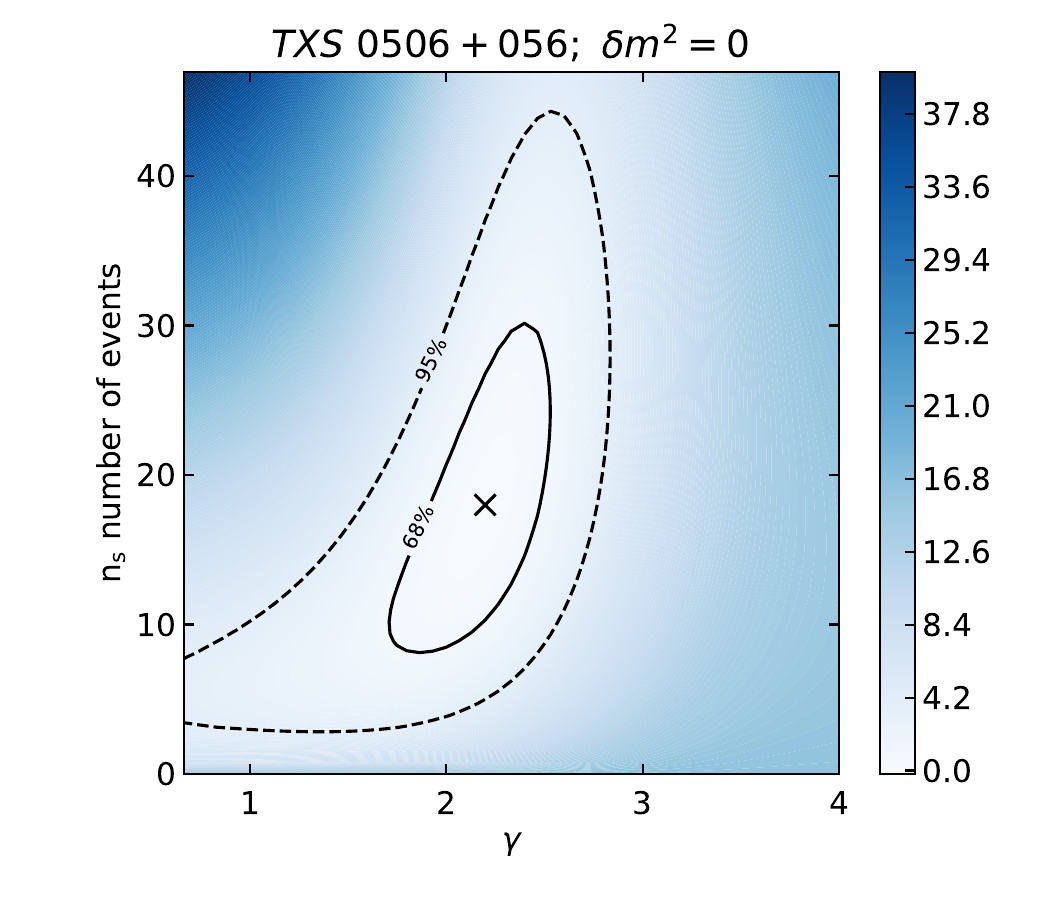}
	\includegraphics[width=.37\textwidth]{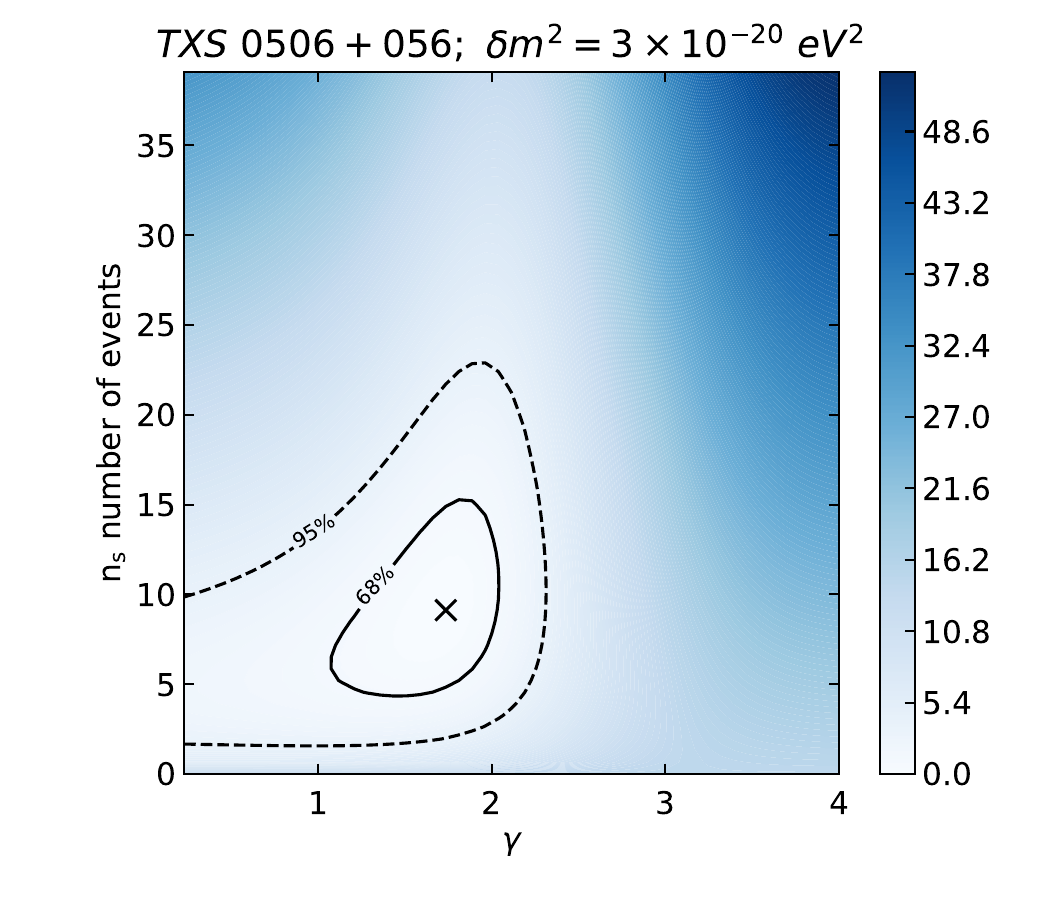}\\
	\includegraphics[width=.37\textwidth]{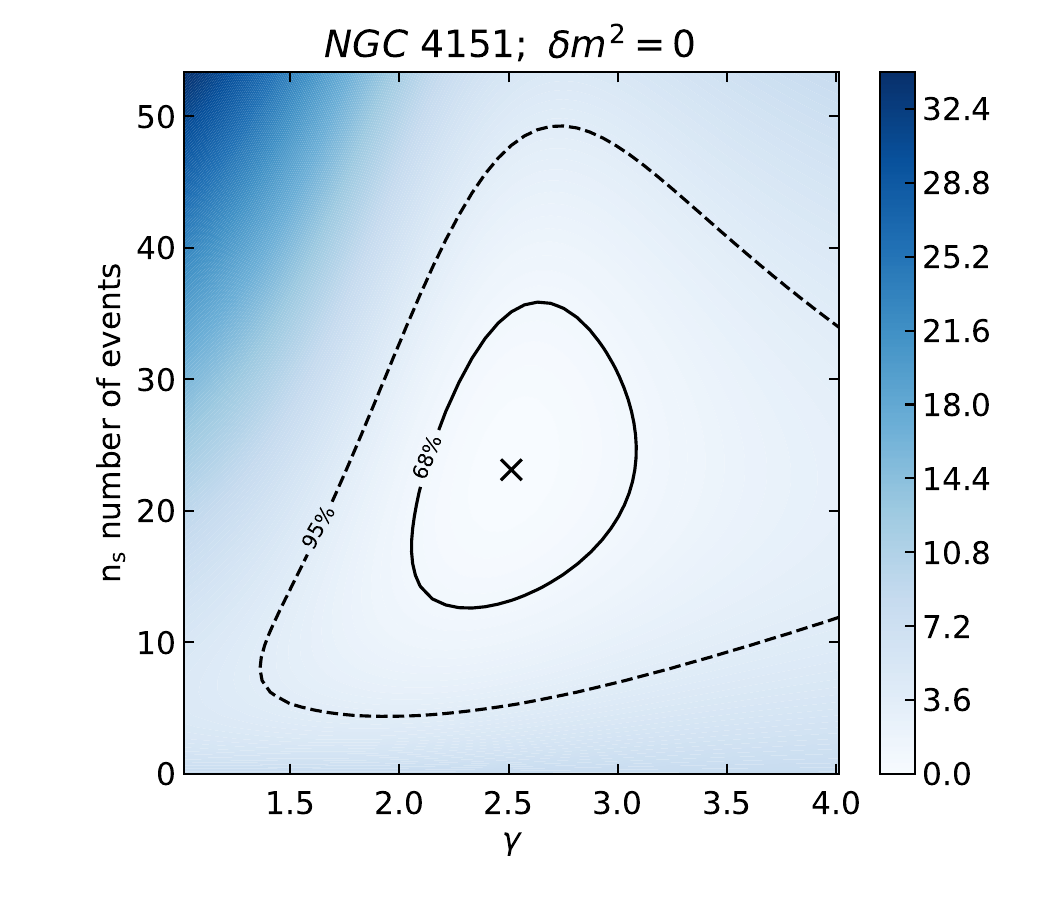}
	\includegraphics[width=.37\textwidth]{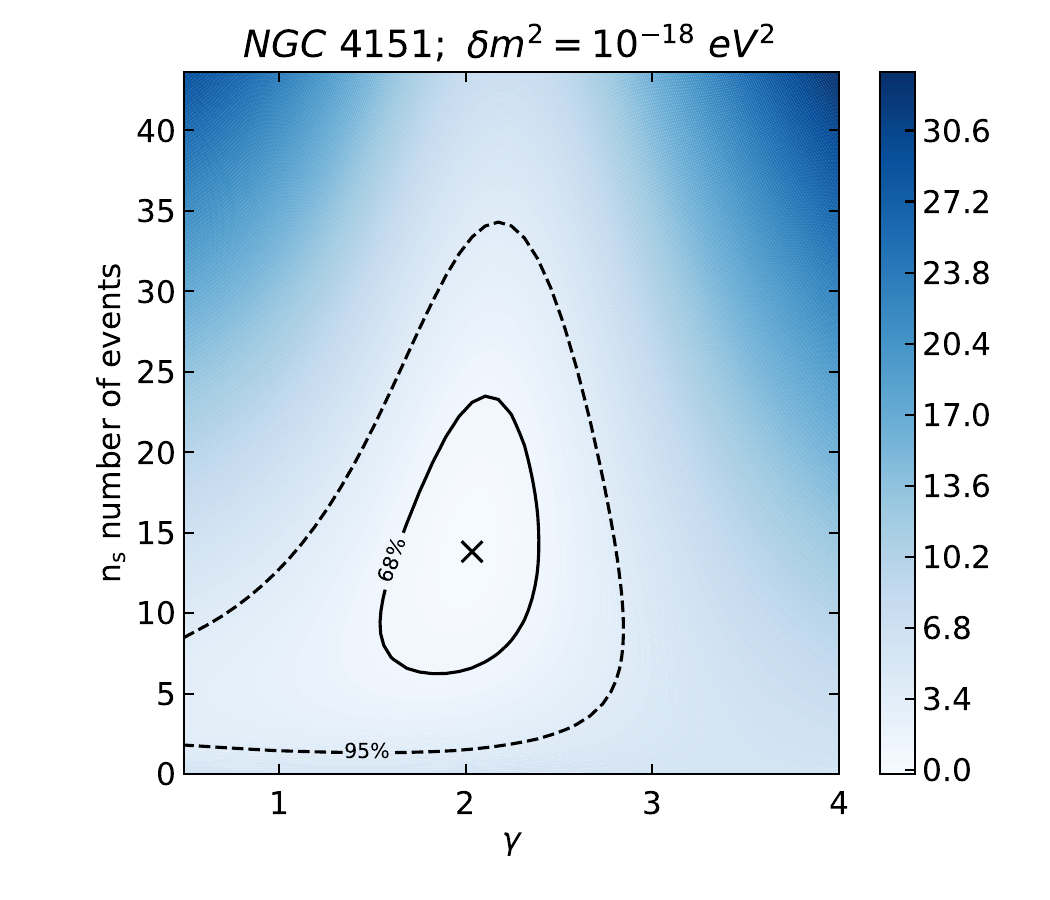}
	\caption{{\bf Data driven background:} Contour plots same as Fig. \ref{fig:contours}. The best-fit point in each case is represented with `x' and the allowed regions with 68\% and 95\% C.L. are shown as solid and dashed curves, respectively. The energy range for IceCube is considered to be 0.5 TeV - 1 PeV.}
	\label{fig:contours_data_driven} 
\end{figure*}

\begin{figure*}[htbp]
\centering	\includegraphics[width=.45\textwidth]{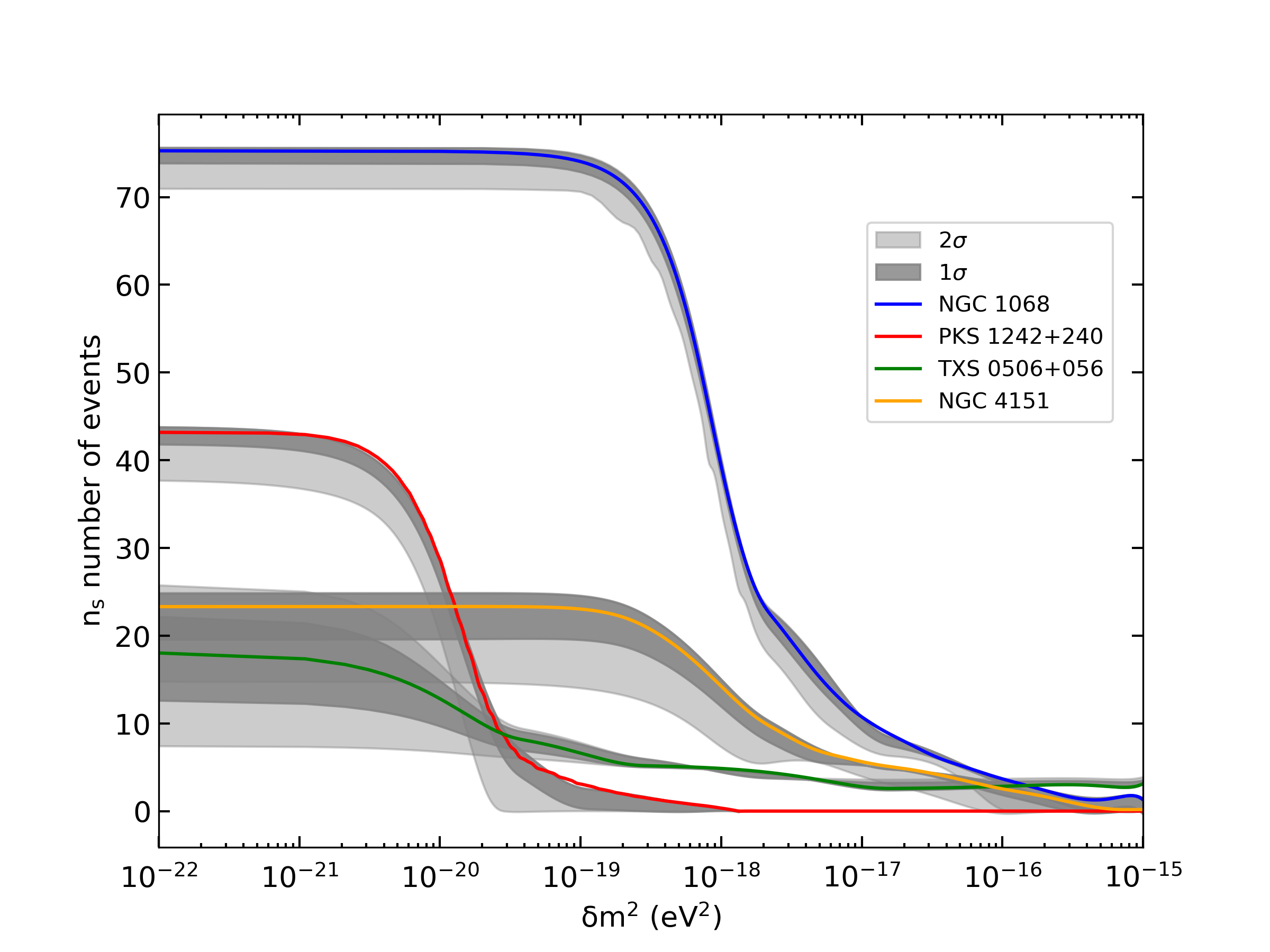}
	\includegraphics[width=.45\textwidth]{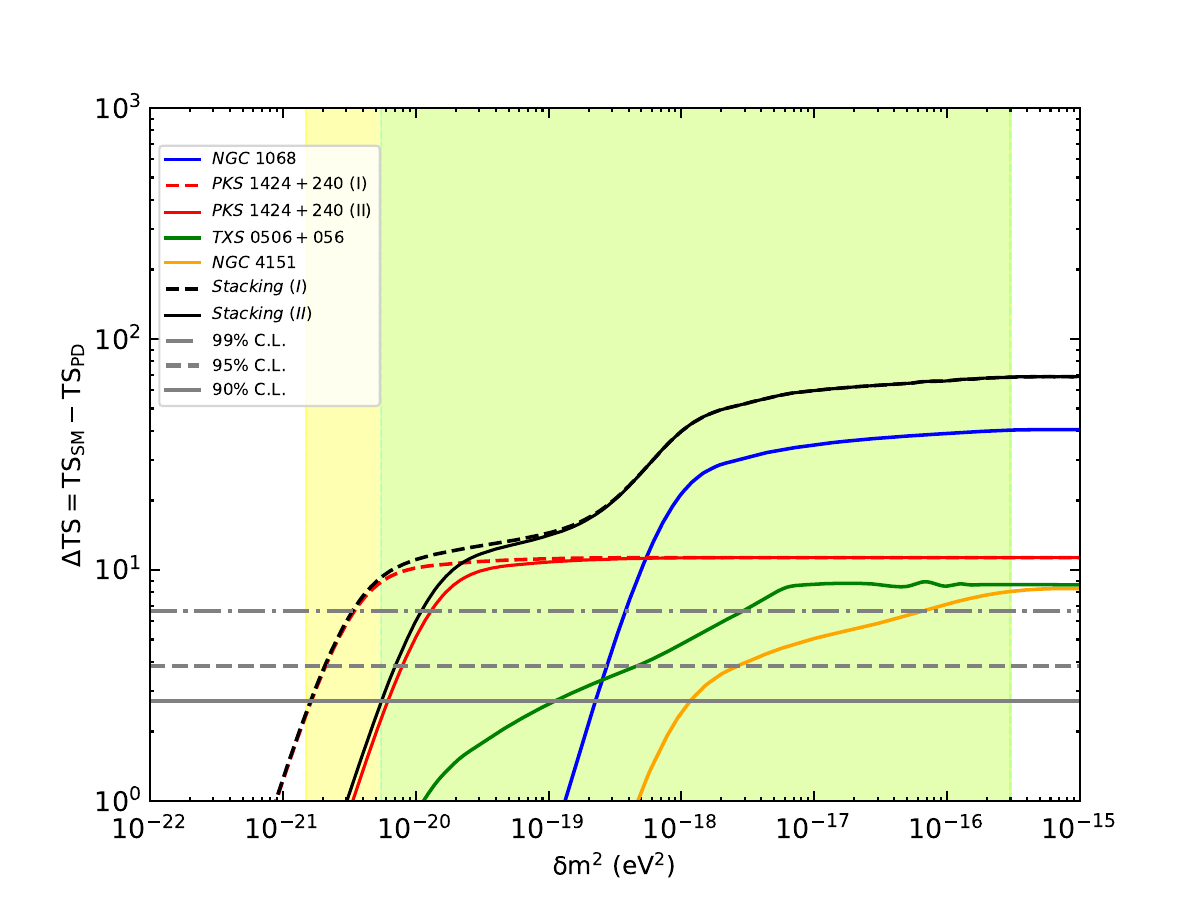}
	\caption{{\bf 0.5 TeV - 1 PeV:} Same as Fig. \ref{fig:stacking} for the case of data-driven background. 
	}
	\label{fig:stacking1} 
\end{figure*}

We show results of our data-driven background analysis in Fig.~\ref{fig:contours_data_driven} both for the standard oscillation scenario ($\delta m^2 = 0$) and for the pseudo-Dirac scenario. We list the best-fit parameter values together with the maximum TS in Table~\ref{tab:best_fit_values} for the standard oscillation scenario. Note that the TS$_{\rm max}$ is decreased in this case for all sources in comparison to the analysis with a fixed background. However, the number of events and the spectral indices for both the backgrounds are still overlapping within the 1$\sigma$ error bars. Figure~\ref{fig:stacking1} shows $n_s$ and $\Delta$TS as functions of $\delta m^2$ in the data-driven background analysis.

We also provide constraints on the $\delta m^2$ parameter for the data-driven background for the neutrino energy interval of 0.5 TeV - 1 PeV which are listed in Table~\ref{tab:results}. Interestingly, the overall exclusion windows for stacking analyses (I) and (II) are notably broader in this case compared to the fixed-background analysis within the same neutrino energy range. 

\section{Discussion and Conclusions}
\label{sec:conclusions}

In this work, we have searched for pseudo-Dirac neutrinos from the most significant four astrophysical point sources in the PSTrack events coming from the IC86 configuration of IceCube given in \cite{IceCube:2021slf} and also including recent detection result of NGC 4151 \cite{IceCube:2024ayt}.  We have fitted these data in the direction of the high energy neutrino sources NGC 1068, PKS 1424+240, TXS 0506+056 and NGC 4151 in the standard three-neutrino oscillation scenario and found good agreement with IceCube Collaboration's published results \cite{IceCube:2022der,IceCube:2024ayt}. 
We have found that our likelihood fits do not improve for any of these sources in the pseudo-Dirac scenario. Hence, we opted for constraining the active-sterile mass splitting $\delta m^2$, which we assumed equal for all three neutrino generations. 

Given varied distances, each source is sensitive above a particular value of $\delta m^2$, below which no constraint can be imposed as there is no difference between the standard and pseudo-Dirac scenarios. Also, there exists an upper bound on the $\delta m^2$ due to detector sensitivity limit as the oscillation length becomes much shorter than the traveled distance and hence the oscillations get averaged out. These limits are reported in Table~\ref{tab:results}. The lower limit of the exclusion window of $\delta m^2$ applies above values of $\delta m_{\rm min}^2$ stated in Table~\ref{tab:results} and up to $\delta m_{\rm max}^2$ at $\ge 90\%$ CL. These limits for NGC 1068, TXS 0506+056 and PKS 1424+240 are in general agreement with the results mentioned in Ref.~\cite{Carloni:2022cqz}.
In the case of NGC 1068, we have found that  
our constraints are stronger than the ones obtained in Ref.~\cite{Rink:2022nvw}, which kept the power-law index of the source flux fixed. Furthermore, only $\nu_\mu + {\bar\nu}_\mu$ source flux and the corresponding survival probability $P_{\mu\mu}$ was used in that analysis. Whereas, we have considered physically motivated pion-decay fluxes of $\nu_\mu + {\bar\nu}_\mu$ and $\nu_e + {\bar\nu}_e$ with corresponding probabilities $P_{\mu\mu}$ and $P_{e\mu}$. 
For individual sources, we have found that PKS 1424+240 gives the lowest value for $\delta m_{\rm min}^2$ whereas NGC 4151 gives the highest value.

We have also performed a stacking analysis by combining the significances of individual sources. These results are also shown in Table~\ref{tab:results}. We found that the stacking analysis significantly broadens the exclusion region for $\delta m^2$ values compared to individual sources, with PKS 1424+240 dominating $\delta m_{\rm min}^2$ and NGC 1068 dominating $\delta m_{\rm max}^2$. These results significantly improve constraints found in previous analyses by other authors.

In our analysis we have varied the neutrino energy in two ranges, 0.1 TeV - 1 PeV and 0.5 TeV - 1 PeV. We found that the exclusion region for $\delta m^2$ broadens slightly when lowering the energy, although the general conclusions made above for individual sources and stacking analysis remain valid.

Additionally, we conducted the same analysis for data-driven backgrounds, focusing on the energy range of 0.5 TeV to 1 PeV for neutrinos. 
We would like to highlight that this is the first time such an analysis has been done using the data-driven background. This analysis resulted in slightly broadening the exclusion window of the active-sterile mass splitting, although the qualitative results and conclusions drawn from the fixed background cases remain valid.

In conclusion, the sensitivity to the pseudo-Dirac scenario for astrophysical sources arises because of a change in the spectral features of the neutrino flux arriving at the earth compared to the standard oscillation. We have not found these signatures of active-sterile neutrino oscillations in the pseudo-Dirac scenario from our analysis of public data release by IceCube. We have independently found exclusion region of the active-sterile mass-squared-difference to be  
$\delta m^2 \in [2.1 \times 10^{-21} - 2.0\times 10^{-16}]$ eV$^2$ at $\ge 90\%$ CL significance from a stacking analysis of four astrophysical neutrino sources in the 0.5~TeV - 1~PeV neutrino energy range. These constraints only weakly depend on the energy range and data-driven background for analysis. Identification of more sources in future may improve this bound further.


\appendix
\section{Appendix}
Here, the event distribution plots for two energy ranges of 0.5 TeV - 1 PeV and 0.1 TeV - 1 PeV are provided. Also, the contour plots, showing the TS values in the $(\gamma - n_s)$-plane just as in Fig. \ref{fig:contours} are given. The results in terms of number of events Vs. $\delta m^2$ and $\Delta$TS Vs. $\delta m^2$ are also presented for 0.1 TeV - 1 PeV energy range.

\begin{figure*}[htbp]
	\centering
	\includegraphics[width=.38\textwidth]{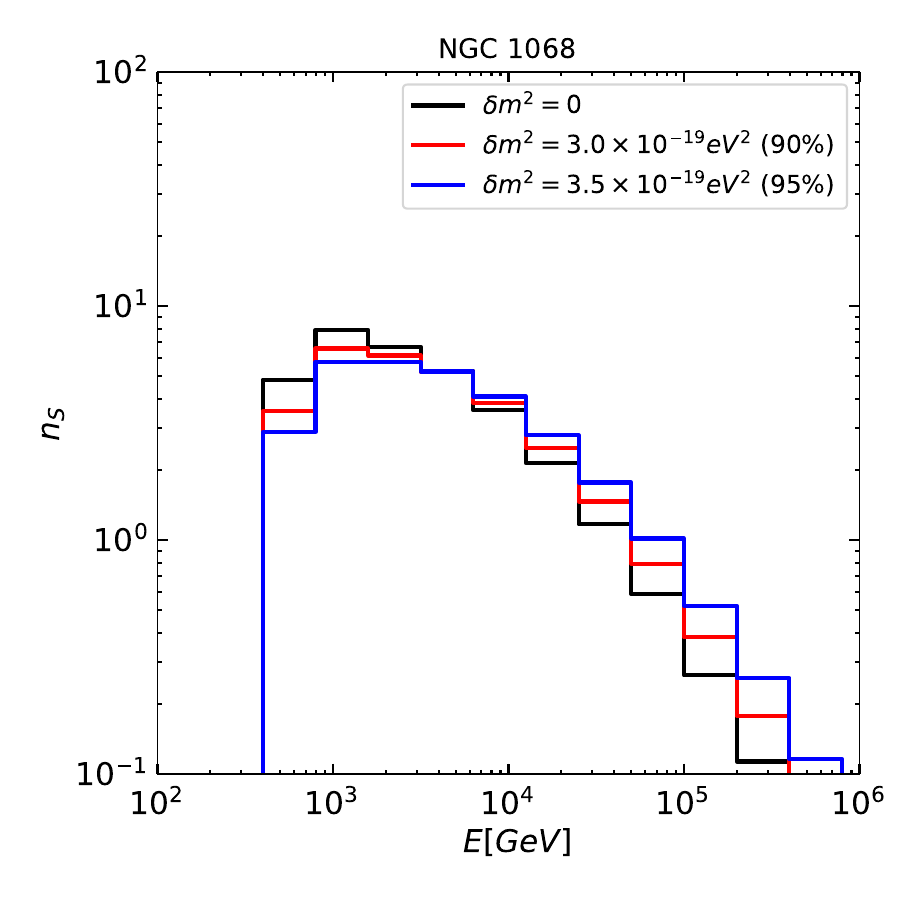}
	\includegraphics[width=.38\textwidth]{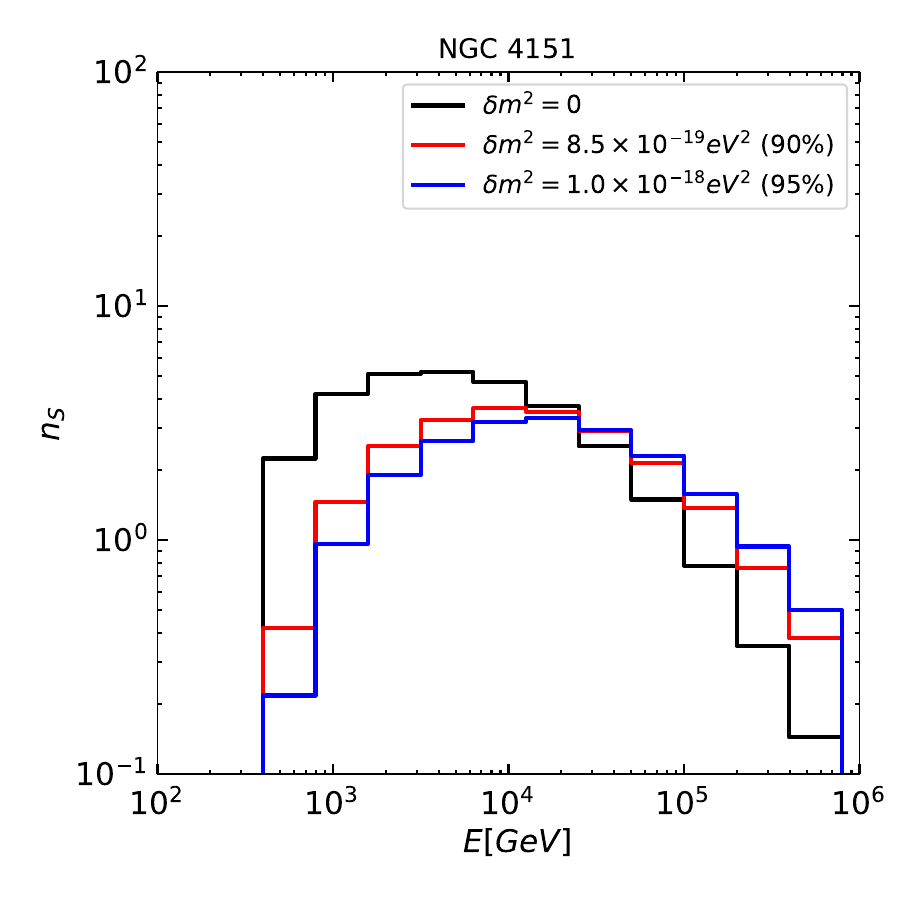}\\
	\includegraphics[width=.38\textwidth]{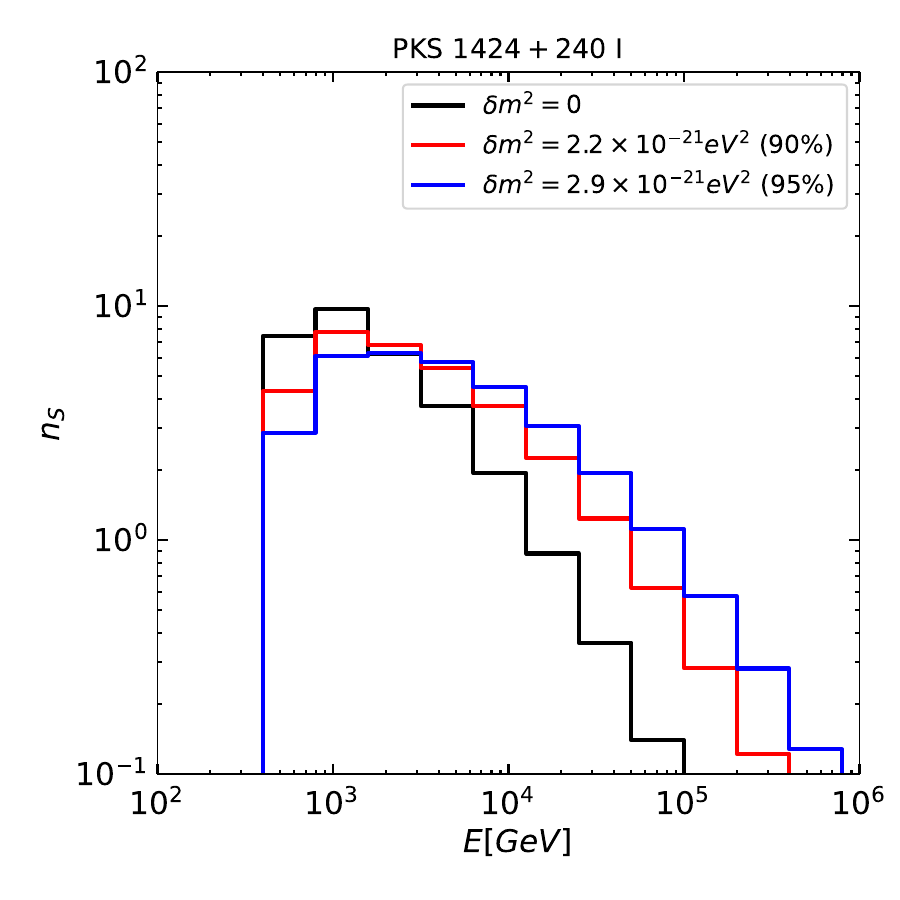}
	\includegraphics[width=.38\textwidth]{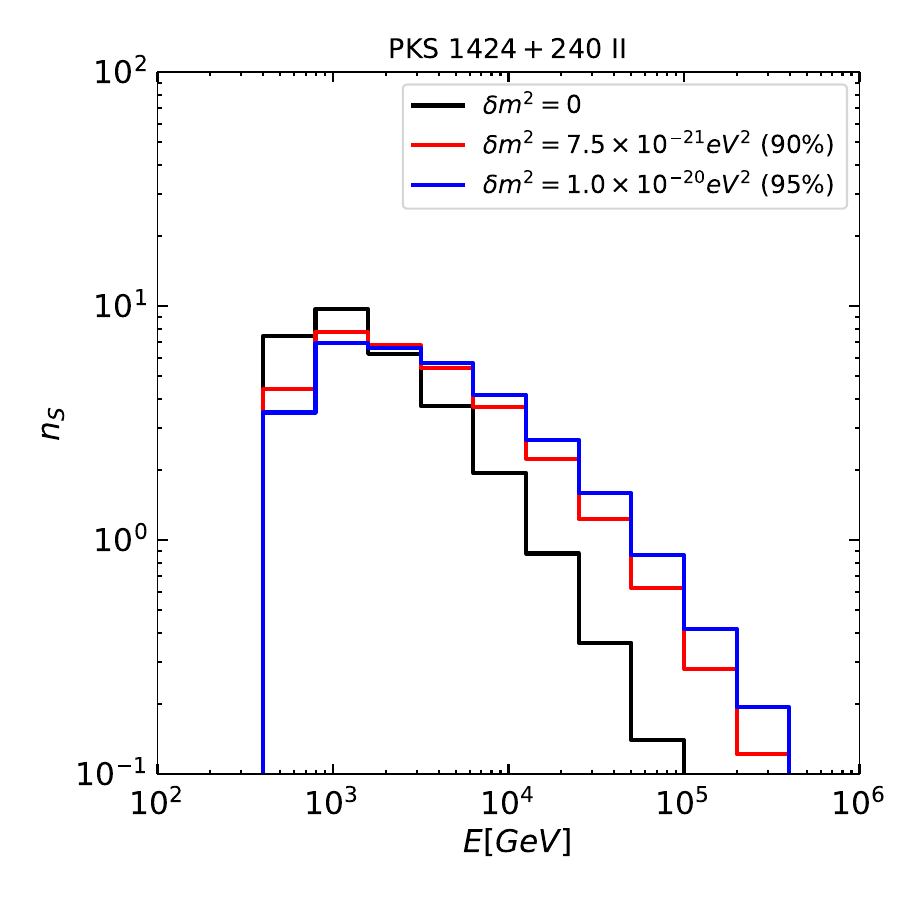}\\
	\includegraphics[width=.38\textwidth]{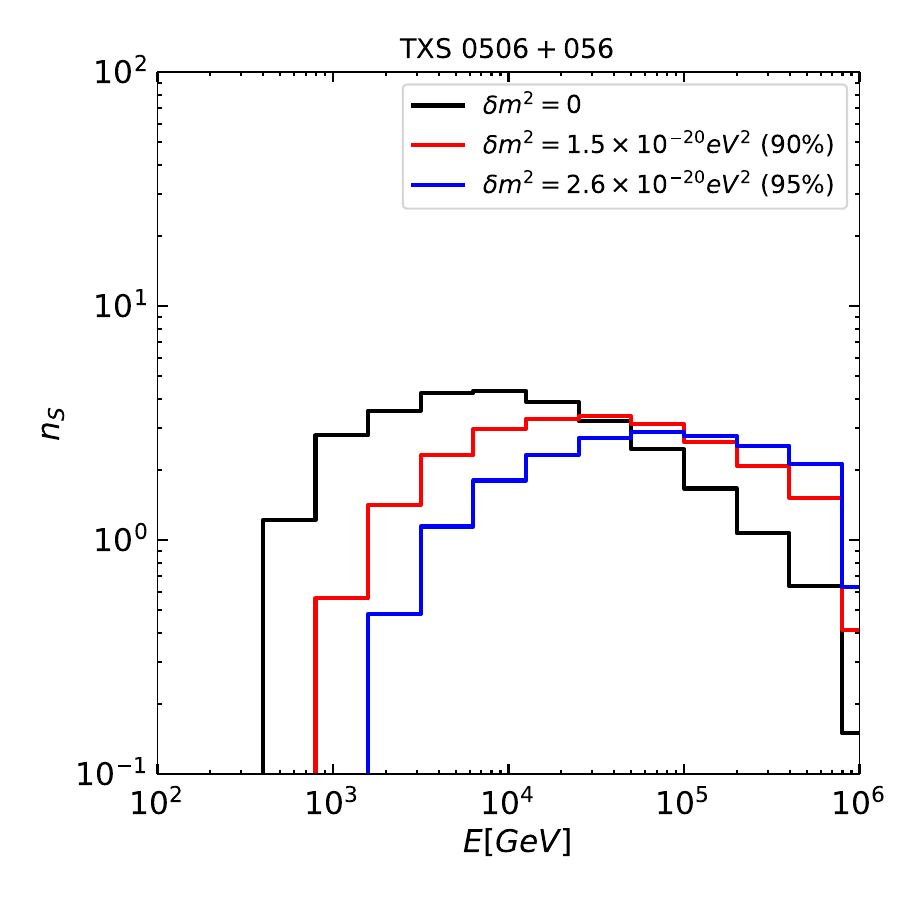}
	\caption{These event distributions are generated for the energy-range 0.5 TeV - 1 PeV for all the sources. The $\hat{\gamma}$ obtained for these fits are 2.79 (2.7) for NGC 1068, 2.0 (1.8) for TXS 0506+056, 2.89 (2.7) for PKS 1424+240 (I) and PKS 1424+240 (II) and 2.0 (1.8) for NGC 4151 for $\delta m^2 \neq 0$ corresponding to 90\% (95\%) CL. The $\gamma_{SM}$ corresponding to the SM scenario are provided in Table~\ref{tab:results}.
	}
	\label{fig:event_0pt5to1000TeV}
\end{figure*}

\begin{figure*}[htbp]
	\centering
	\includegraphics[width=.4\textwidth]{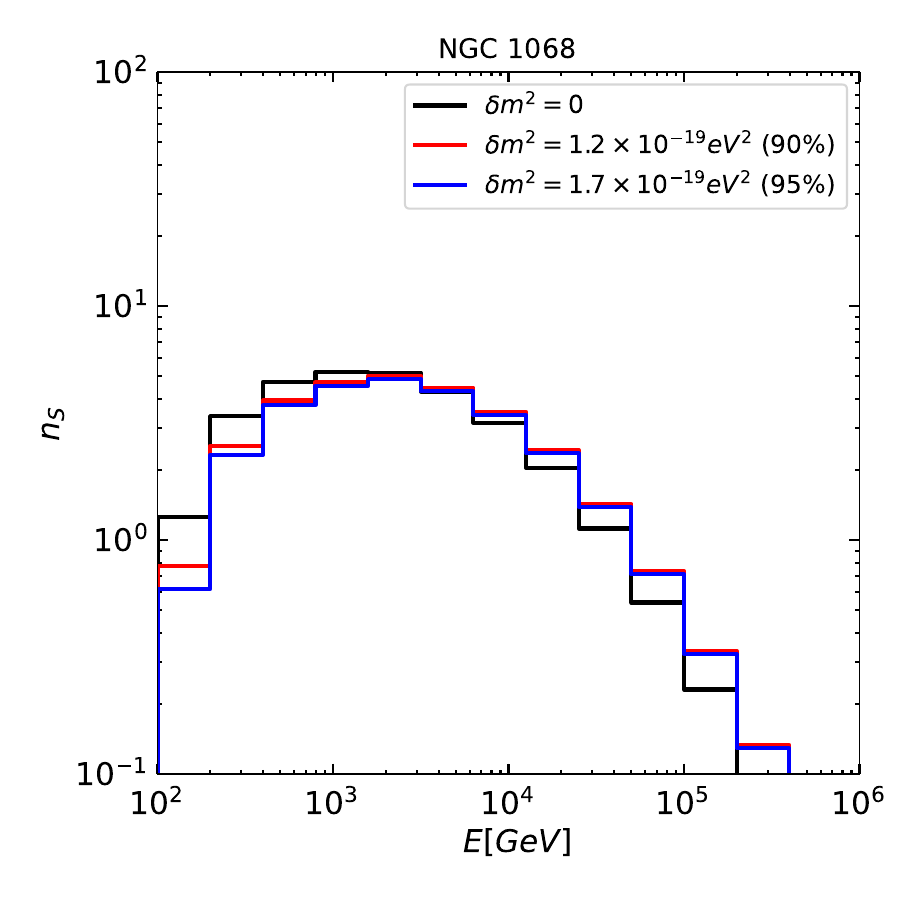}
	\includegraphics[width=.4\textwidth]{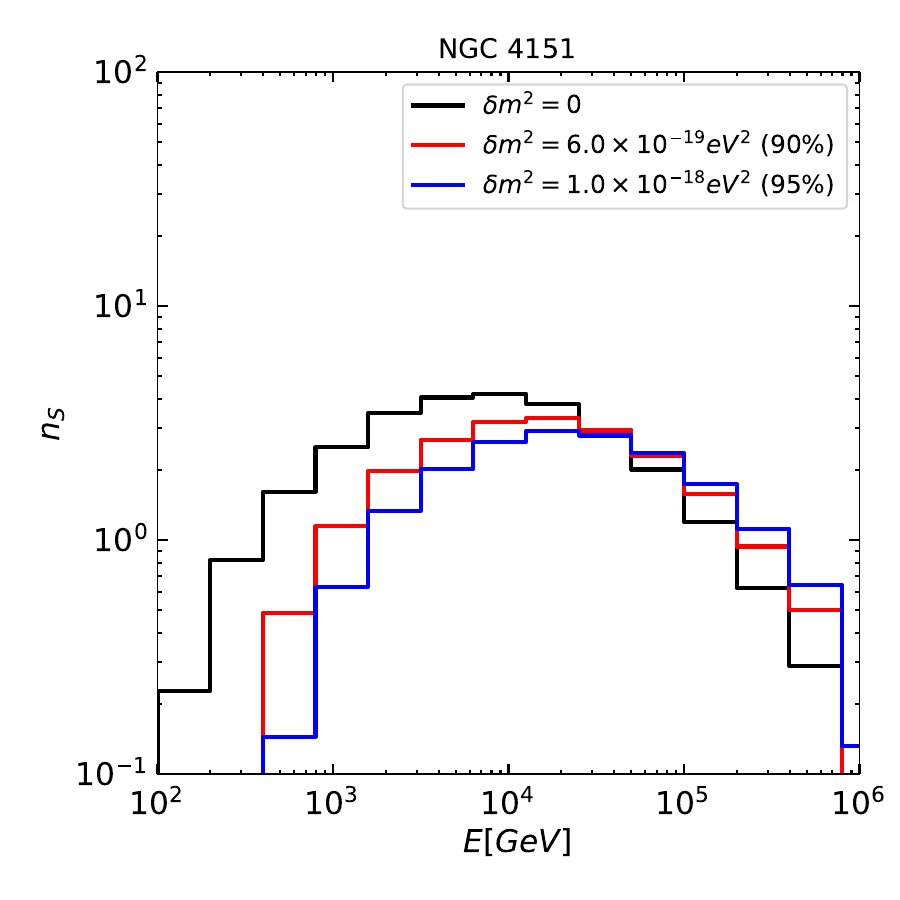}\\
	\includegraphics[width=.4\textwidth]{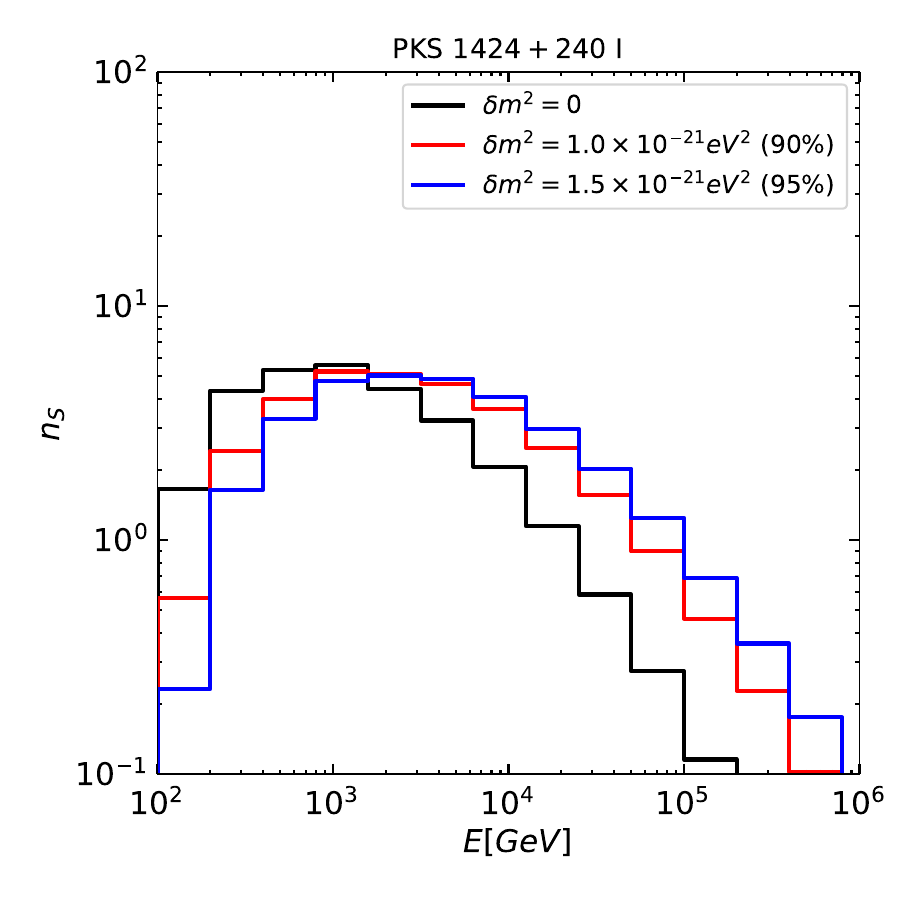}
	\includegraphics[width=.4\textwidth]{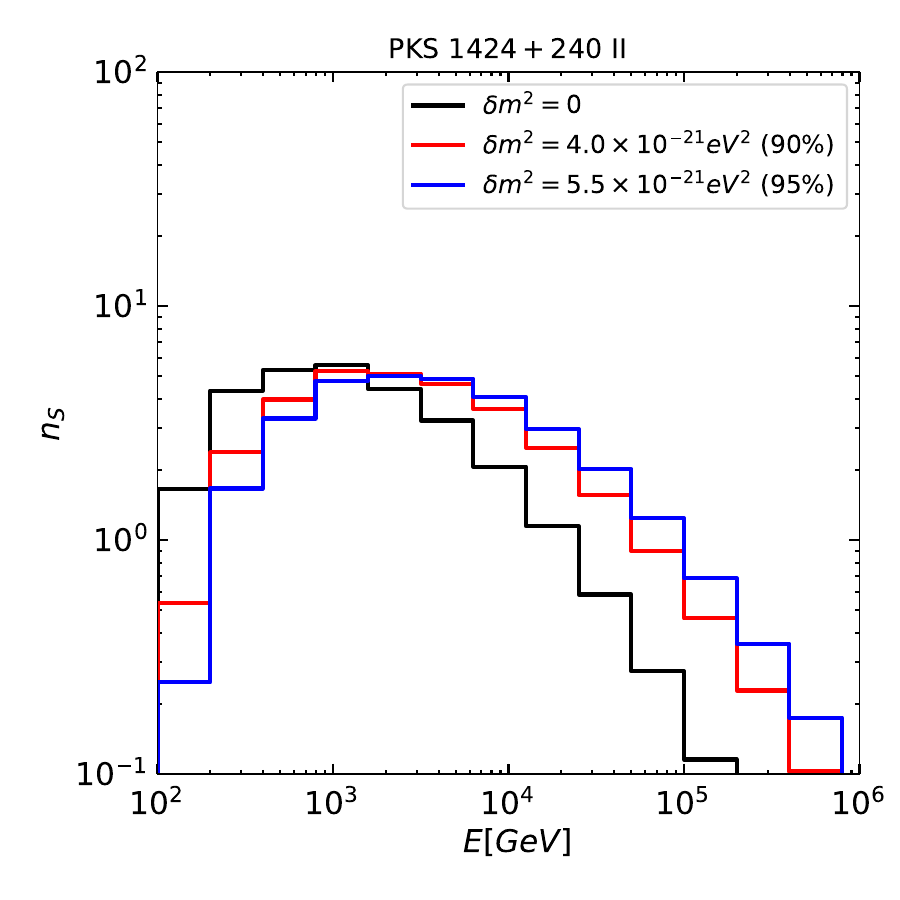}\\
	\includegraphics[width=.4\textwidth]{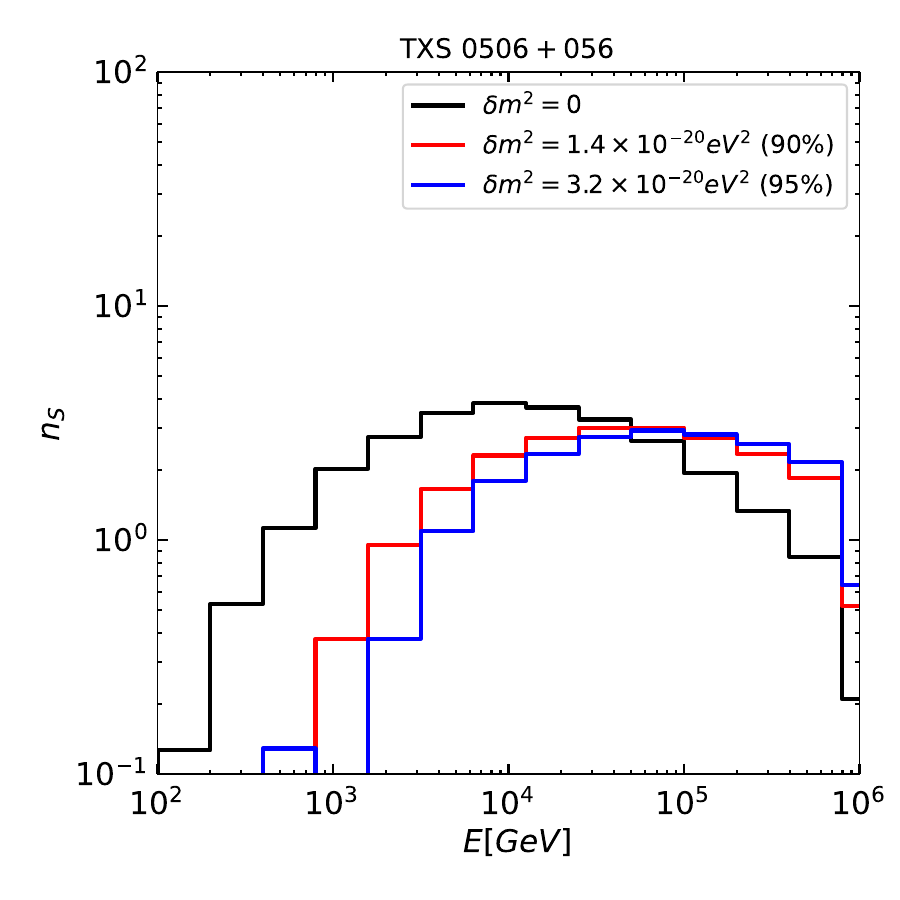}
	\caption{Same as Fig. \ref{fig:event_0pt5to1000TeV} for 0.1 TeV - 1 PeV energy range. The $\hat{\gamma}$ obtained are 2.7 (2.6) for NGC 1068, 1.89 (1.69) for TXS 0506+056, 2.7 (2.6) for PKS 1424+240 (I) and PKS 1424+240 (II) and 2.0 (1.9) for NGC 4151 with 90\% (95\%) CL constraint on $\delta m^2$.
	}
	\label{fig:event} 
\end{figure*}

\begin{figure*}[htbp]
	\centering
	\includegraphics[width=.37\textwidth]{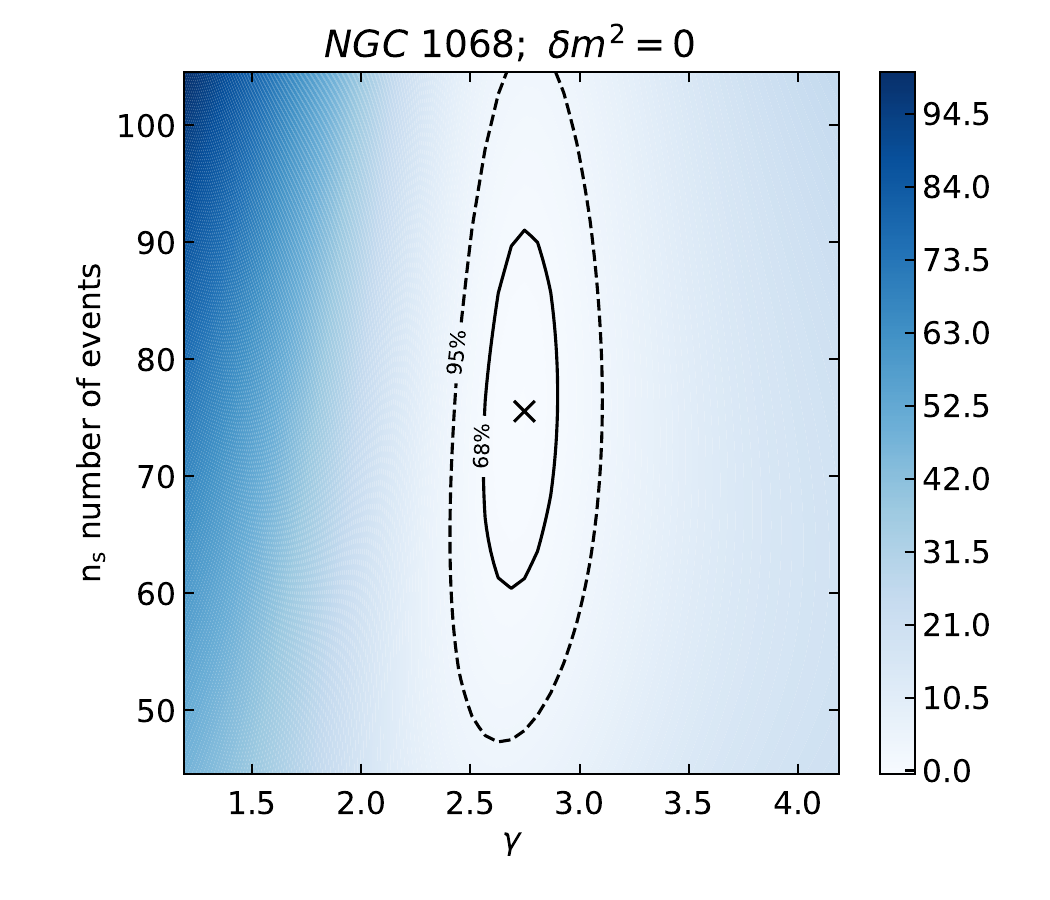}
	\includegraphics[width=.37\textwidth]{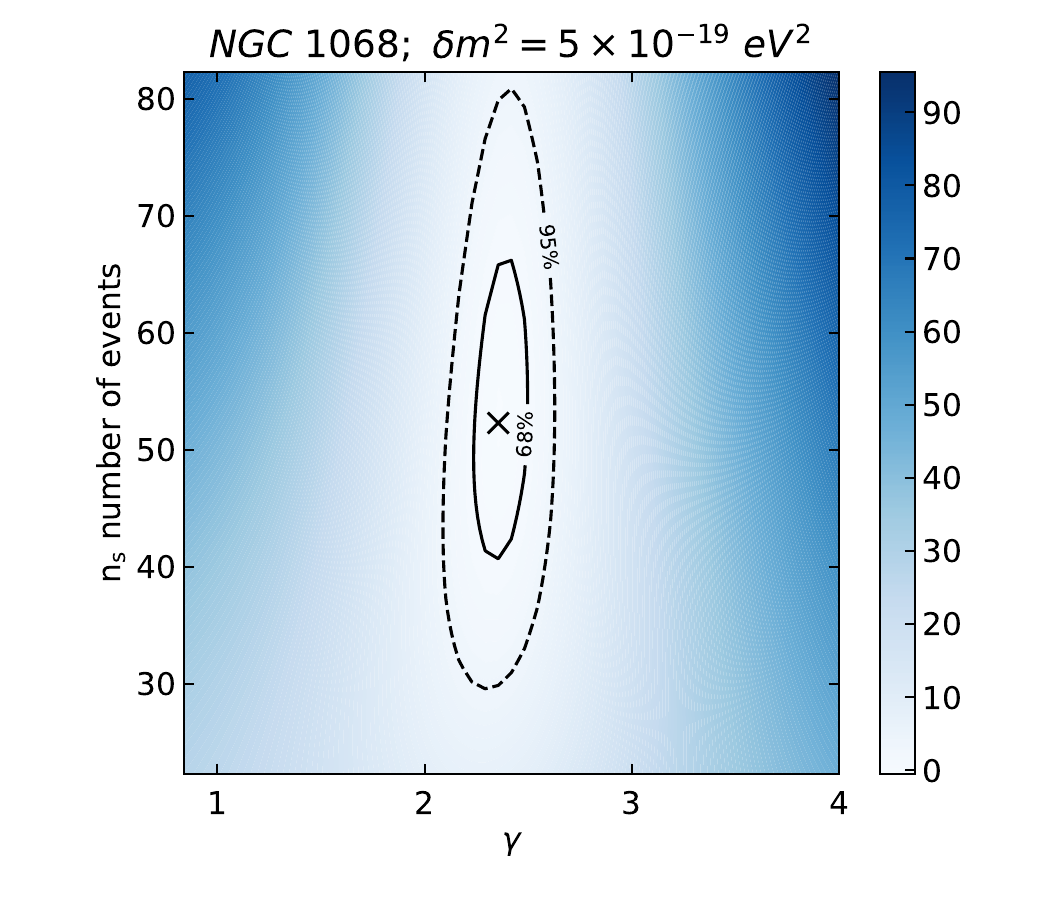}\\
	\includegraphics[width=.37\textwidth]{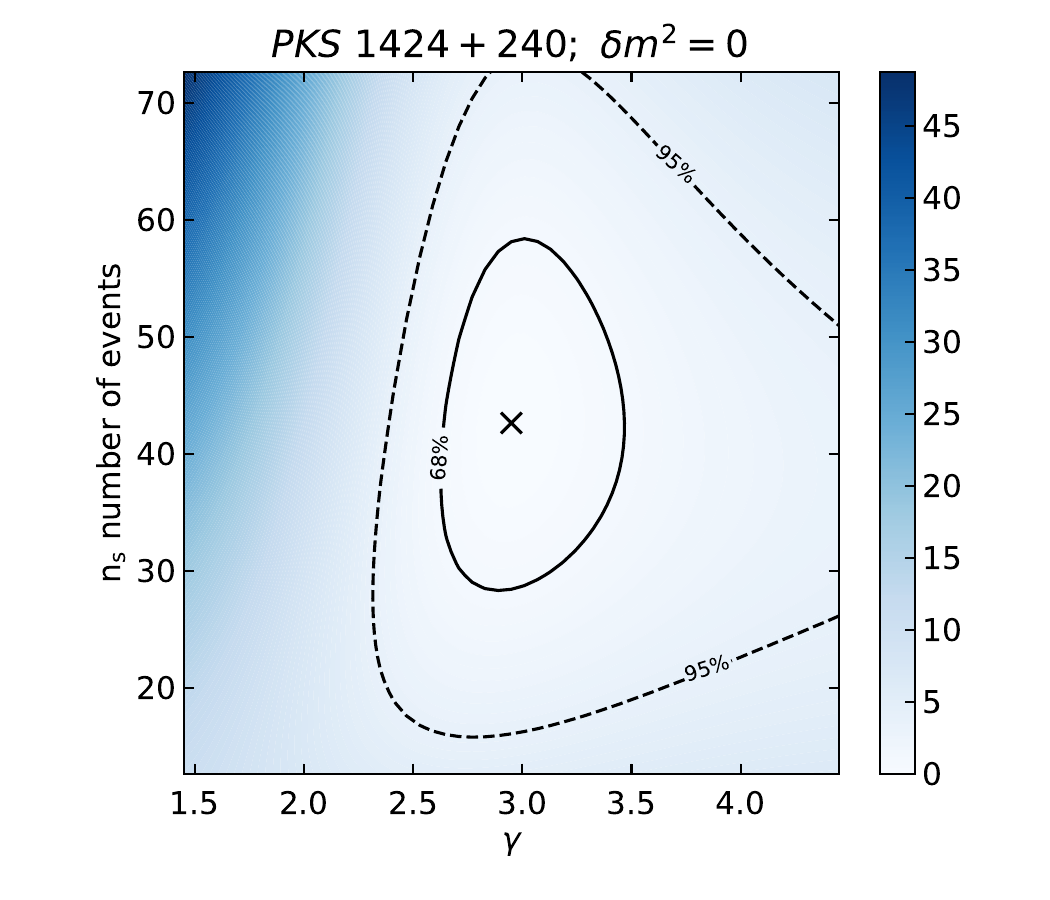}
	\includegraphics[width=.37\textwidth]{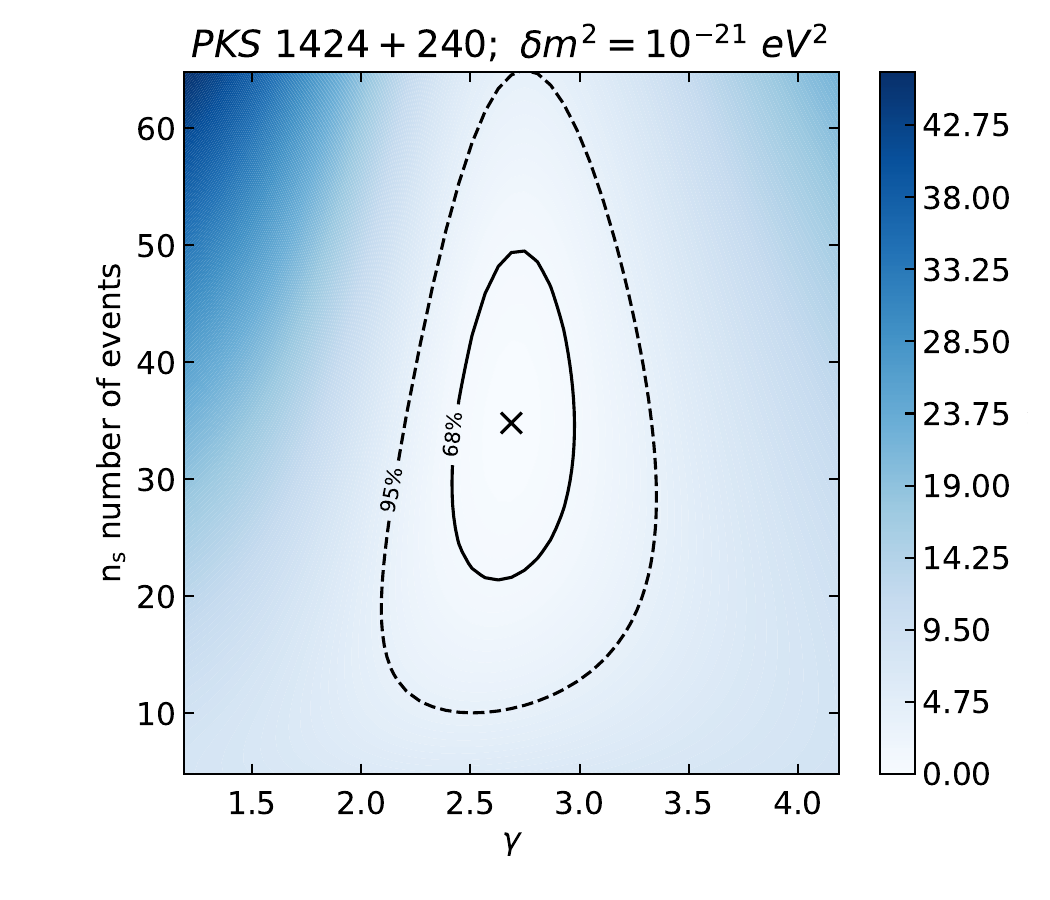}\\
	\includegraphics[width=.37\textwidth]{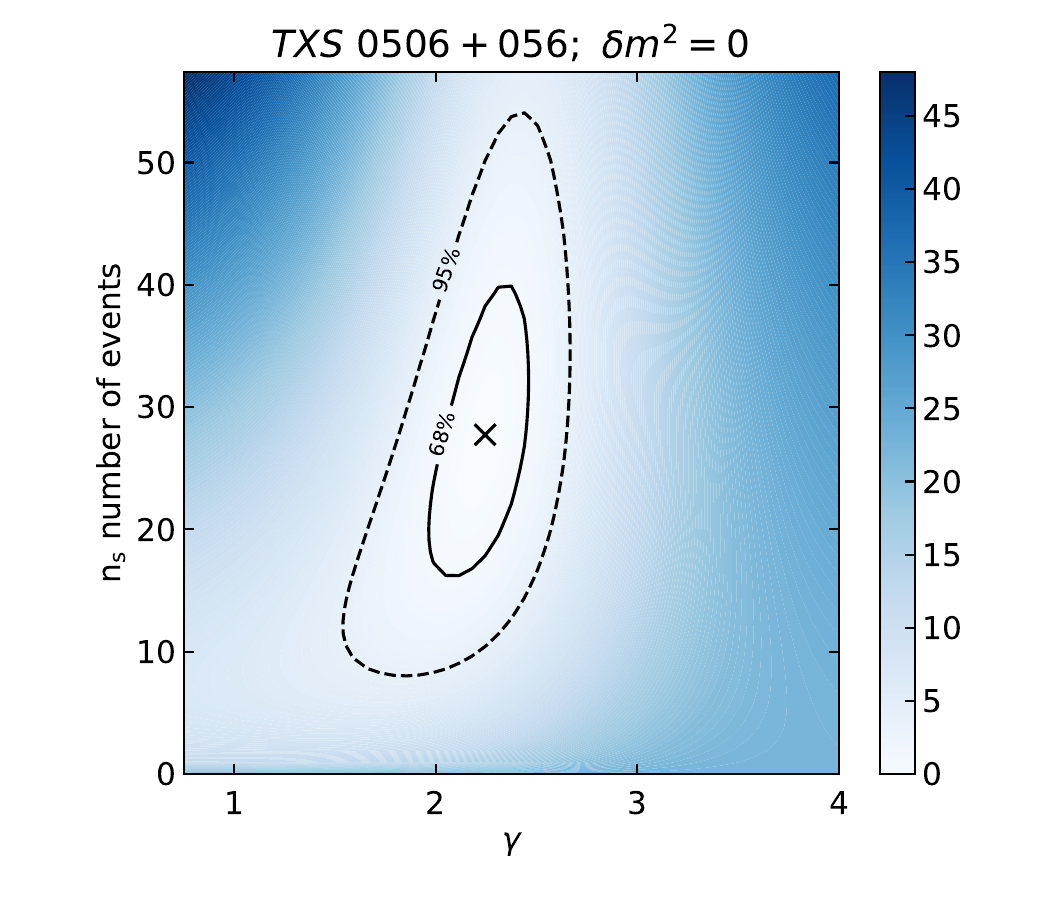}
	\includegraphics[width=.37\textwidth]{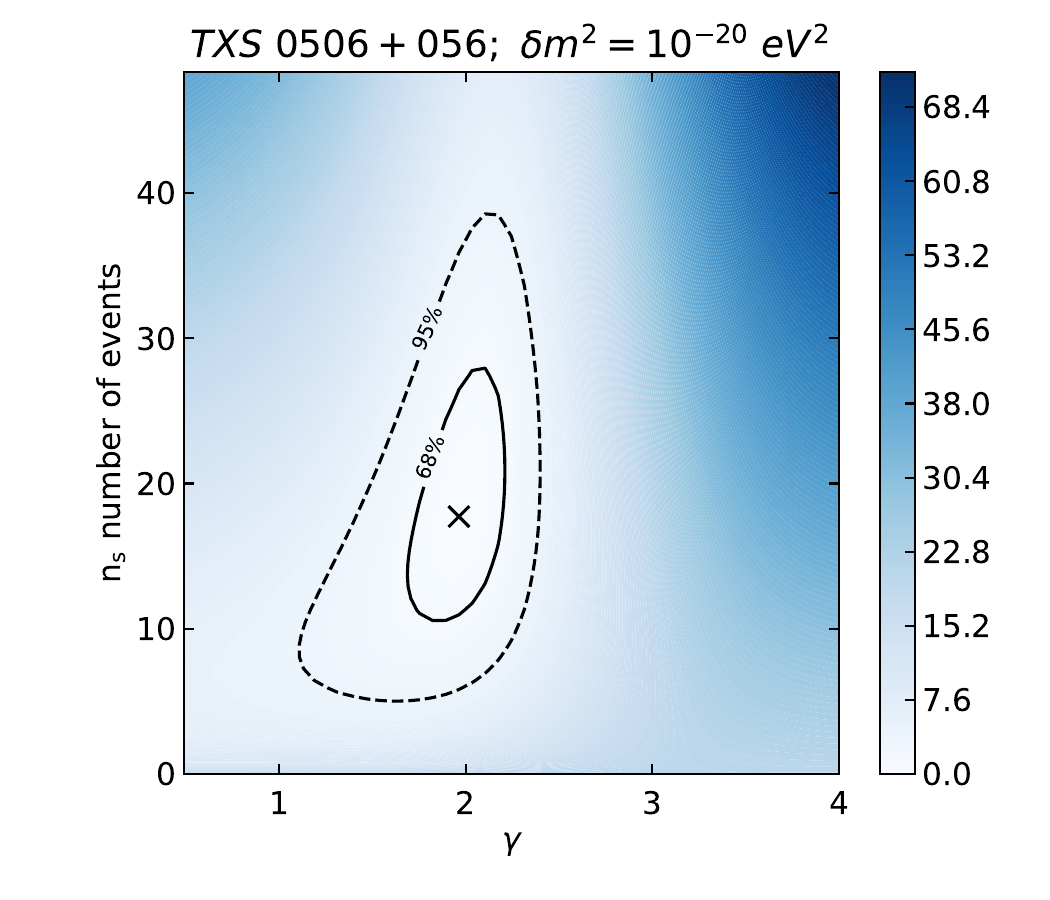}\\
	\includegraphics[width=.37\textwidth]{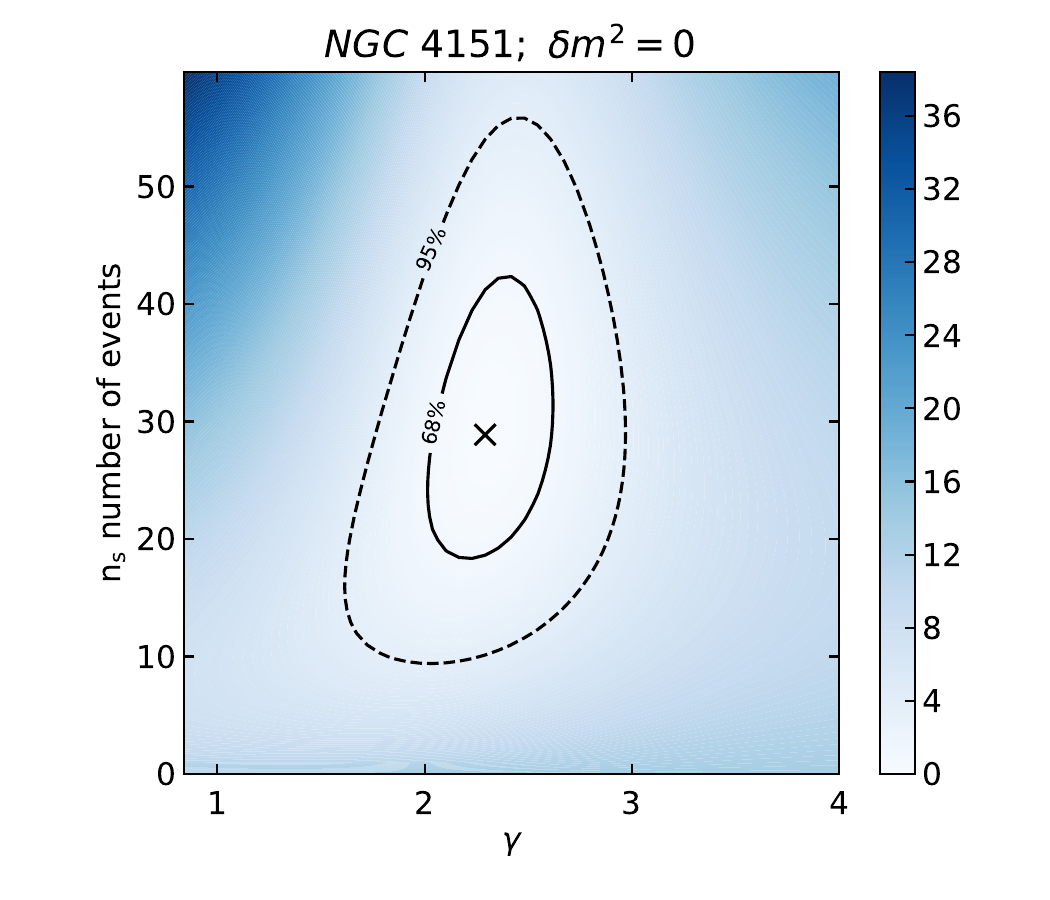}
	\includegraphics[width=.37\textwidth]{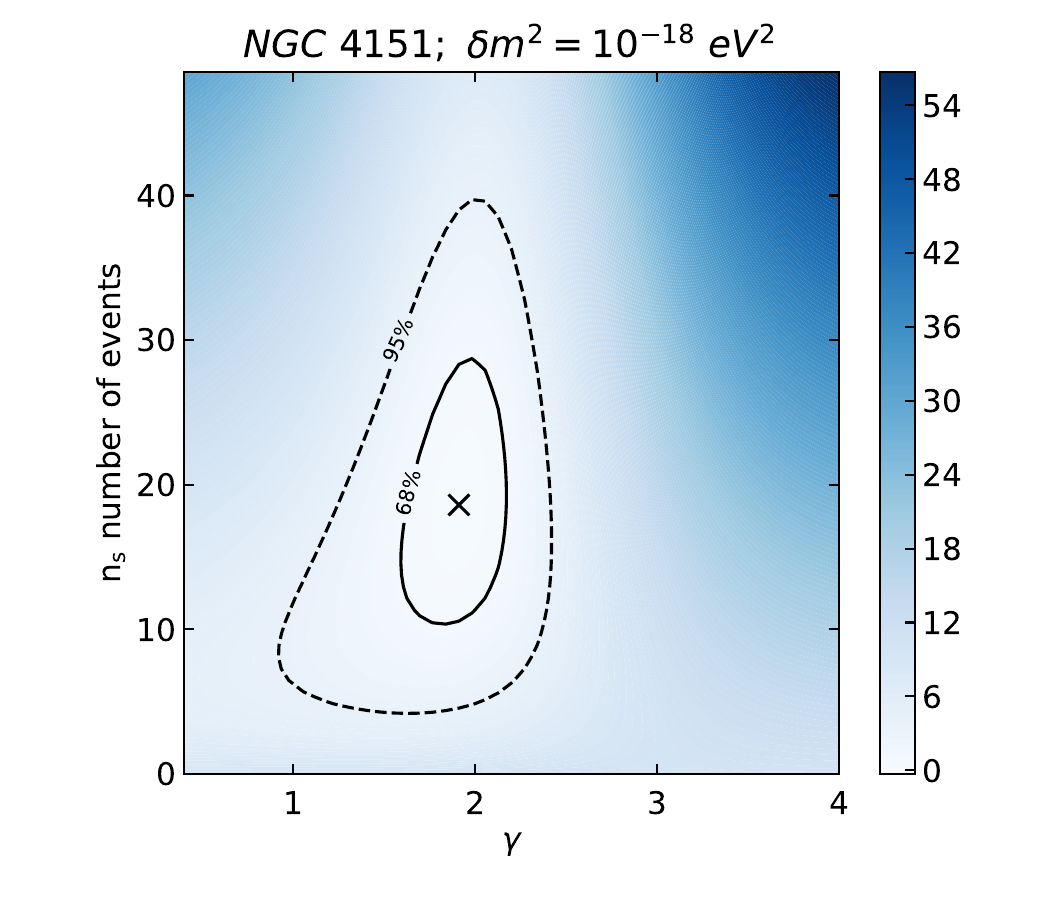}
	\caption{{\bf 0.1 TeV - 1 PeV:} Same as Fig. \ref{fig:contours}.
	}
	\label{fig:contours_0pt1to1000TeV} 
\end{figure*}

\begin{figure*}[htbp]
\centering	\includegraphics[width=.45\textwidth]{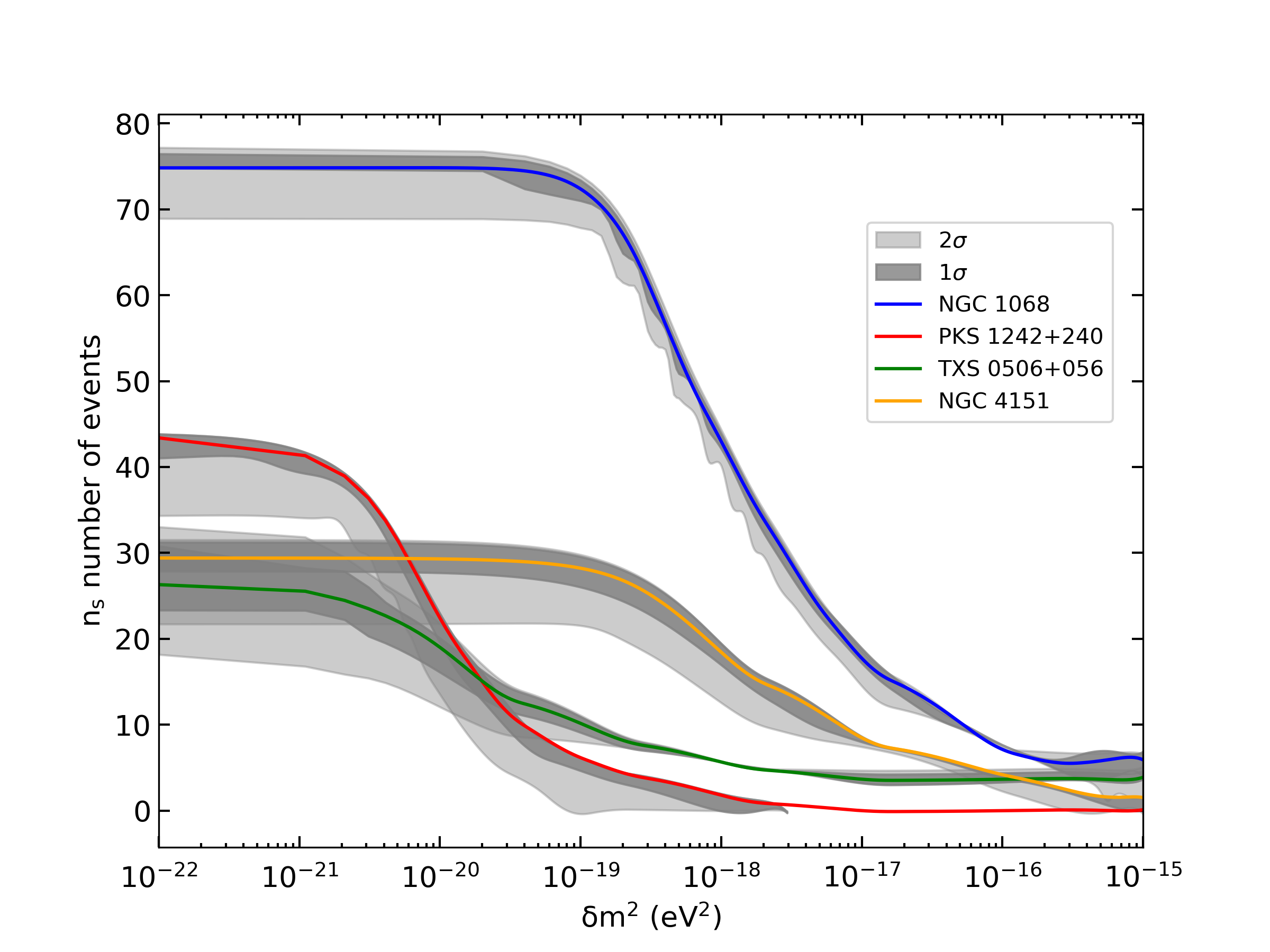}
	\includegraphics[width=.45\textwidth]{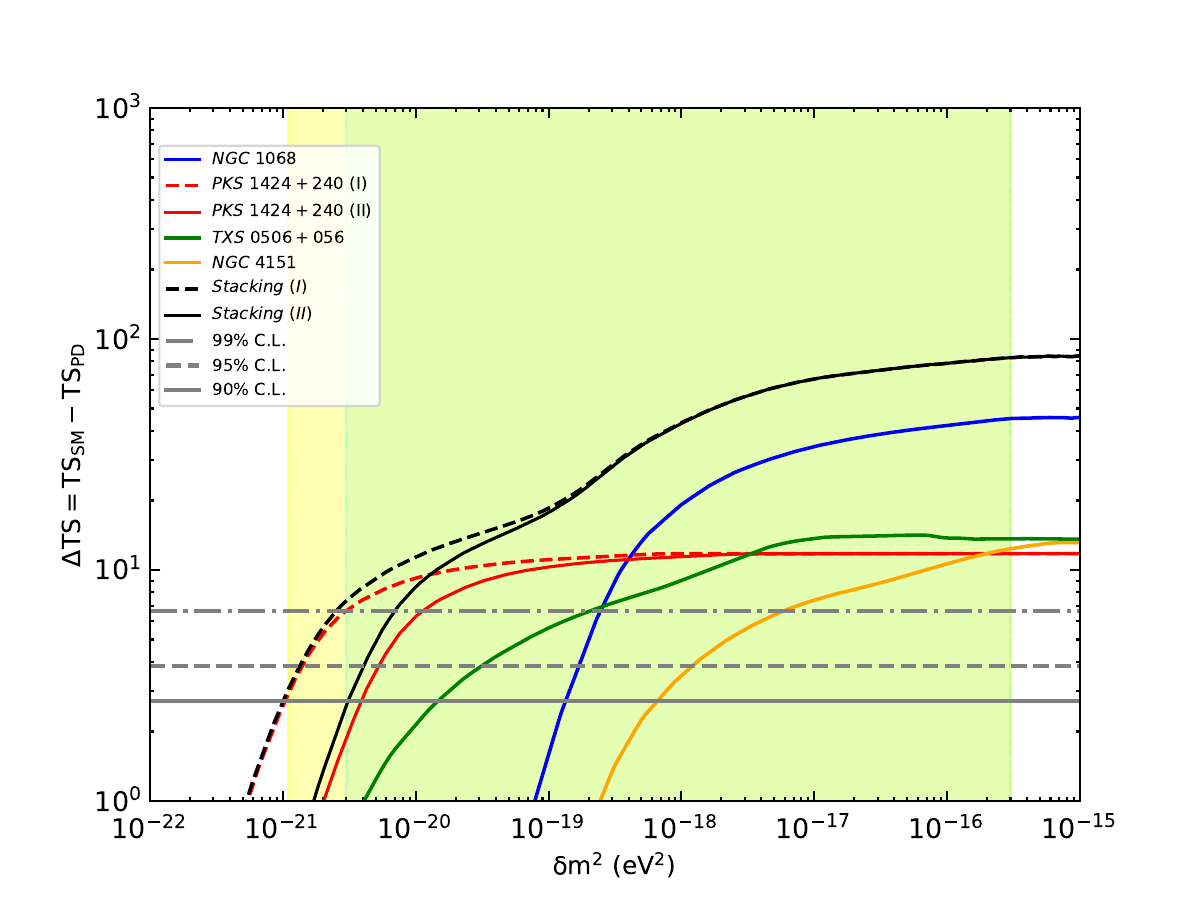}
	\caption{{\bf 0.1 TeV - 1 PeV:} Same as Fig. \ref{fig:stacking}   
	}
	\label{fig:stacking2} 
\end{figure*}
\FloatBarrier

\begin{acknowledgements}
We thank  Carlos~A.~Arguelles and P.~S. Bhupal Dev for useful discussions. This work was partially supported by grants from the National Research Foundation (NRF), South Africa, through the National Institute of Theoretical and Computational Sciences (NITheCS) and from the University of Johannesburg Research Council. L. S. M. also thanks IFSC/USP - Bolsa PRPI USP no. 22.1.08498.01.0 and FAPESP grants: 2023/12705-0, 2019/ 14893-3 for financial support.
\end{acknowledgements}



\end{document}